\documentclass[fleqn,usenatbib]{mnras}

\usepackage{newtxtext,newtxmath}

\usepackage[T1]{fontenc}
\usepackage{ae,aecompl}

\usepackage [english]{babel}
\usepackage [autostyle, english = american]{csquotes}
\MakeOuterQuote{"}

\usepackage{graphicx}	
\usepackage{amsmath}	
\usepackage{amssymb}	

\usepackage[normalem]{ulem}

\title[SGRB environments]{Constraints on the circumburst environments of short gamma-ray bursts}

\author[O'Connor et al. (2020)]{
Brendan O'Connor$^{1,2,3,4}$\thanks{E-mail: oconnorb@gwu.edu},
Paz Beniamini$^{5,1,2}$, and 
Chryssa Kouveliotou$^{1,2}$
\\
$^{1}$Department of Physics, The George Washington University, 725 21st Street NW, Washington, DC 20052, USA \\
$^{2}$Astronomy, Physics, and Statistics Institute of Sciences (APSIS), The George Washington University, Washington, DC 20052, USA \\
$^{3}$Department of Astronomy, University of Maryland, College Park, MD 20742-4111, USA \\
$^{4}$Astrophysics Science 
Division, NASA Goddard Space Flight Center, 8800 Greenbelt Rd, Greenbelt, MD 20771, USA\\
$^{5}$Division of Physics, Mathematics and Astronomy, California Institute of Technology, Pasadena, CA 91125, US
}

\date{Accepted XXX. Received YYY; in original form ZZZ}

\pubyear{2020}

\begin{document}
\label{firstpage}
\pagerange{\pageref{firstpage}--\pageref{lastpage}}
\maketitle

\begin{abstract}
Observational follow-up of well localized short gamma-ray bursts (SGRBs) has left $20-30\%$ of the population without a coincident host galaxy association to deep optical and NIR limits ($\gtrsim 26$ mag). These SGRBs have been classified as observationally hostless due to their lack of strong host associations. It has been argued that these hostless SGRBs could be an indication of the large distances traversed by the binary neutron star system (due to natal kicks) between its formation and its merger (leading to a SGRB). The distances of GRBs from their host galaxies can be indirectly probed by the surrounding circumburst densities. We show that a lower limit on those densities can be obtained from early afterglow lightcurves. We find that $\lesssim16\%$ of short GRBs in our sample took place at densities $\lesssim10^{-4}$ cm$^{-3}$. These densities represent the expected range of values at distances greater than the host galaxy's virial radii. We find that out of the five SGRBs in our sample that have been found to be observationally hostless, none are consistent with having occurred beyond the virial radius of their birth galaxies. This implies one of two scenarios. Either these observationally hostless SGRBs occurred outside of the half-light radius of their host galaxy, but well within the galactic halo, or in host galaxies at moderate to high redshifts ($z\gtrsim 2$) that were missed by follow-up observations.
\end{abstract}




\section{Introduction}

Short gamma-ray bursts (SGRBs) are bright flashes of gamma-rays with a typical duration $\lesssim2$ s \citep{Chryssa93}. The recent simultaneous detection of a binary neutron star (BNS) merger by the LIGO-Virgo Collaboration \citep[GW 170817/GRB 170817A;][]{Abbott2017}, the {\it Fermi}/Gamma Ray Burst Monitor \citep[GBM;][]{Goldstein2017}, and the INTernational Gamma-ray Astrophysics Laboratory \citep[INTEGRAL;][]{Savchenko2017} confirmed the association of SGRBs with BNS mergers. Extensive follow up observations localized the event, GW 170817, to a position within the half-light radius of its host galaxy \citep[NGC4993;][]{Coulter2017,Levan2017,Troja2017}. In contrast to this event, $20-30\%$ of the well localized population of SGRBs have been found to lack a strong host galaxy association, and therefore have been classified as hostless, hereafter referred to as \textit{observationally} hostless \citep{Berger2010a,Fong2013, FongBerger2013, Tunnicliffe2014}. Follow-up with the Hubble Space Telescope (HST) of these observationally hostless SGRBs has set limits of 26 mag on host galaxy detection \citep{Perley2009,FongBergerFox,Rowlinson2010,Berger2010a,Fong2012,Fong2013,FongBerger2013,Tunnicliffe2014}.
\par
The connection between a GRB and its host galaxy has been traditionally established either through probabilistic arguments \citep[e.g.,][]{Bloom2002,Bloom2007,Levan2007,Campisi2008,Berger2010a} or by the superposition of the GRB localization region with a galaxy \citep[e.g.,][]{Cobb2006,Cobb2008}. However, unless the redshift of the GRB is spectroscopically measured from its afterglow lines, which is not typically the case for SGRBs (as they are not bright enough), it is not possible to rule out the possibility that the galaxy superposition is due to random sky alignment with either a foreground or background galaxy. Moreover, it is often the case that the SGRB lacks a coincident galaxy but is surrounded by a number of plausible hosts. This potentially leads to multiple galaxies with a similar probability of random alignment with the SGRB localization region \citep{Berger2010a,Fong2013,FongBerger2013,Tunnicliffe2014}. The probability of random alignment, also know as probability of chance coincidence $P_{cc}$, can be estimated by calculating the probability to identify a galaxy of equal magnitude or brighter within the localization region \citep[e.g.,][]{Bloom2002,Bloom2007,Levan2007,Berger2010a,Blanchard2016}. If this probability is too high (or equivalent for multiple galaxies in the field) the GRB is deemed to be observationally hostless.

\par
 The preferred explanation for SGRBs that lack a strong host association is that prior to merger the BNS system has escaped the host galaxy due to the change in the center-of-mass velocity of the system following the second supernova \citep{Lyne1994,Hansen1997,Fryer1999,Wex2000,Belczynski2006, Church2011,Behroozi2014,Zevin2019}. The mass ejection in the second supernova as well the explosion asymmetry induce a kick on the newly formed BNS. This kick can increase the center-of-mass velocity of the system eventually allowing it to potentially escape its host galaxy due to large merger delay times of $10^7-10^{11}$ years \citep{Piran1992,Zheng2007,Zemp2009,Behroozi2014}. \citet{Behroozi2014} found that natal kicks on the order of $180$ km/s can explain the population of observationally hostless SGRBs, which have large offsets (30-100 kpc) from their lowest $P_{cc}$ potential host \citep{Berger2010a}. In contrast, using observations of Galactic BNS systems, \citet{Beniamini2016} and \citet{Beniamini2016p2} found that the majority ($60-70\%$) of their natal kick velocities are $\sim 30$ km/s. It has also recently been shown that a significant fraction ($\gtrsim 40-60\%$) of Galactic BNS merge rapidly on scales less than 1 Gyr \citep{Beniamini2019}. Low kick velocities and shorter time delays lead to a small fraction\footnote{The escape velocity from the center of a typical SGRB host with virial mass $M_{200}=10^{12}\,M_\odot$ is $450$ km/s \citep{NFW1997,Bullock2001,Behroozi2014}. The fraction of BNS receiving $v_\textrm{CM}>450$ km/s is $\lesssim 0.5\%$ \citep{Beniamini2016}.} of BNS escaping their galactic halo and thus becoming physically hostless SGRBs. We have defined here a \textit{physically} hostless GRB as having occurred outside of its birth galaxy's virial radius. 
\par
In this work, we use the observed circumburst density distribution of SGRBs to infer their physical offset distribution and at the same time determine the consistency of this distribution with the observed offsets from putative SGRB host galaxies. We apply the standard GRB afterglow model \citep{Meszaros1997,Sari1998} to early afterglow observations from the \textit{Neil Gehrels Swift Observatory} (hereafter \textit{Swift}) X-ray Telescope (XRT) in order to obtain lower limits on circumburst densities. These limits allow us to determine the maximum fraction of SGRBs occurring below a given density threshold, $f(<n)$. We find the fraction of SGRBs that occur at $n\approx 10^{-6}\,\textrm{cm}^{-3}$ to be $\lesssim 2\%$ (i.e., $f(<10^{-6})\lesssim 2\%$). Adopting a typical density at the virial radius of $10^{-4}$ cm$^{-3}$ (see \S \ref{sec: gas density profile}), we obtain a physically hostless fraction of $f(<10^{-4})\lesssim16\%$. We derive circumburst density limits for five observationally hostless SGRBs, and find that none of these limits are lower than $10^{-4}$ cm$^{-3}$. It is likely, therefore, that the observationally hostless fraction of SGRBs is dominated by events that do occur within their host's virial radius. Furthermore, our methods can be used to exclude potential host galaxies as the true birth galaxy by considering whether the inferred density for an observed offset is consistent with these lower limits. It is important to understand the impact of false host galaxy associations on the physical properties of any given sample of SGRBs. An incorrect host galaxy identification can bias the observed offset and redshift distributions of SGRBs which in turn affects our understanding of their luminosity function, natal kicks, delay times, rates, and likely environments. In an upcoming publication we will address the fraction of SGRBs expected to occur at a given host offset using forward modeling of BNS formation and kinematics within their birth galaxies to constrain the intrinsic observationally hostless fraction. 
\par
We outline the paper as follows. In \S \ref{sec: obs}, we  present the sample criteria and observations. We then describe our method of calculating lower limits on the circumburst density in  \S \ref{sec: methods}. We present results in \S \ref{sec_Results} and our conclusions in \S \ref{sec: conclusions}. Throughout the work, we apply a flat cosmology with parameters $H_o=67.4$ and $\Omega_\Lambda=0.685$ \citep{Planck2018} and adopt the convention that $\log$ represents $\log_{10}.$

\section{Observations}
\label{sec: obs}
\subsection{Sample}
\subsubsection{SGRBs}
\label{subsec: SGRBs Sample}
We use the \textit{Swift}/XRT GRB Lightcurve Repository\footnote{https://www.swift.ac.uk/xrt\_curves/} to obtain afterglow observations of 52 SGRBs \citep{Evans2007,Evans2009}. We have chosen to incorporate SGRBs with $T_{90}\lesssim 0.8\,$s from the \textit{Swift} online database\footnote{https://swift.gsfc.nasa.gov/archive/grb\_table/} through June 2019. We impose this duration requirement to minimize contamination from a possible collapsar (long gamma-ray burst, hereafter LGRB) population \citep{Bromberg2012,Bromberg2013}. We also include SGRBs from \citet{Berger2013}, \citet{Fong2015}, and \citet{Wanderman2015} (hereafter, WP15) with $T_{90}$ above this cutoff value (but $\lesssim 2$\,s).
\par

We define the afterglow as radiation emitted from the forward shock, and we have, therefore, chosen to exclude bursts with potential contamination from the prompt emission \citep[e.g., high-latitude emission (HLE),][]{KumarPanaitescu2000,Nousek2006}. If the X-ray lightcurve shows an early-steep decline (e.g., $F_X\propto t^{-\alpha}$ with $\alpha\gtrsim1.5$) or an internal plateau\footnote{GRBs with external plateaus are not excluded from this analysis, such as those observed in, e.g., GRBs 090510, 101219A, 130603B, and 140903A.} \citep[e.g., with $\Delta t/t\ll 1$; see for example,][]{Zhang2006,Liang2006,Troja2007}, we define the afterglow onset as the first detection after this period, which follows the ubiquitous GRB afterglow lightcurve decay (e.g., $\alpha\sim1$). We likewise exclude SGRBs with only a single afterglow detection (see \S \ref{sec: completeness}). This is a conservative approach as in this work we are interested in only upper limits on the afterglow peak time $t_p$ and lower limits on peak X-ray flux $F_{X,p}$; under the assumption the emission is due to the forward shock. The peak of the afterglow lightcurve is also the onset of deceleration, which occurs when the blast-wave has piled up a significant enough amount of ambient medium that the bulk Lorentz factor, $\Gamma$, begins to decrease substantially; therefore, the deceleration time is $t_p=R_p/2c\Gamma^2$, where $R_p$ is the deceleration radius \citep{Blandford1976}. We define the limits on the deceleration time and peak flux as $t_o$ and $F_{X,o}$ respectively, which represent the time and corresponding flux of the first clear XRT afterglow detection for each burst. The X-ray flux is obtained in the standard \textit{Swift} $0.3-10$ keV band. We collect this information for each burst in Table \ref{Table_SGRB_DATA}.
\par
We correct the observed flux for absorption with the Portable, Interactive Multi-Mission Software\footnote{https://heasarc.gsfc.nasa.gov/docs/software/tools/pimms.html} (PIMMS) using the early-time X-ray photon index and hydrogen column density $N_H$ (Galactic and intrinsic) from the \textit{Swift} data reduction pipeline. In this sample, only 19 (36\%) SGRBs have measured redshifts. Redshifts were collected from \citet{Fong2015}, \citet{Berger2013}, and WP15 (and references therein). For the cases of GRB 150423A and GRB 160821B, we adopt the redshifts from \citet{Malesani2015} and \citet{Levan2016GCN} (see also \citep{Xu2016GCN,Lamb2019,Troja2019}) respectively.
\par
We use the Band function \citep{Band1993} fits to the time-averaged spectra of the BAT data, published in the \textit{Swift} Repository, to compile the $\gamma$-ray fluence, $\phi_\gamma$, in the $15-150$ keV band. The Band function fit is also used to determine the bolometric correction (see equation (\ref{eqn: kbol}) below) for this sample of SGRBs, which is used to determine the true emitted energy of the burst. For the low-energy spectral index, $\alpha$, and observer frame peak $\gamma$-ray energy, $E_{p,\textrm{obs}}$, we use the average values of short GRBs from \textit{Fermi}/GBM detections \citep{Nava2011}. These are  $\log(E_{p,\textrm{obs}}/\mbox{keV})=2.69\pm0.19$ and $\alpha=-0.5\pm 0.4$. We also adopt a high-energy spectral index of $\beta=-2.25$, following WP15. The high-energy spectral index is of little importance for this work because when $\beta<-2$ the bolometric correction is largely determined by the low-energy spectral index and the peak energy. We convert the average peak energy in the observer frame $E_{p,\textrm{obs}}$, taken from \citet{Nava2011}, to the source frame using the average redshift for our sample $\langle z\rangle =0.84$ which yields $E_{p,\textrm{source}}=900$ keV. We note $\langle z\rangle =0.84$ is the mean redshift for bursts in our sample that have measured redshift ($36\%$), and that it is quite likely the other $54\%$ of bursts have preferentially higher redshifts \citep[see e.g.,][]{Ghirlanda2016}. Our results are consistent with a similar analysis previously applied by WP15, who used a $\langle z\rangle =0.69$. 

We adopt $E_{p,\textrm{source}}=900$ keV in our calculations of the bolometric correction factor for those bursts without measurements of $E_{p,\textrm{obs}}$. For bursts with \textit{Fermi}/GBM data, we use the measured values of $E_{p,\textrm{obs}}$, $\alpha$ and $\beta$. The bolometric correction $k_{\textrm{bol}}$ is 
calculated using 
\begin{align}
k_{\textrm{bol}}=\frac{\int^{10\,\textrm{MeV}}_{1\,\textrm{keV}}N(E)\,E\,dE}{\int^{(1+z)\,150\,\textrm{keV}}_{(1+z)\,15\,\textrm{keV}}N(E)\,E\,dE}
\label{eqn: kbol}
\end{align}
where $N(E)$ is the energy spectrum of the GRB prompt emission. In this paper, we adopt a  Band function for $N(E)$ \citep{Band1993} with $\alpha$ and $\beta$ discussed above. We denote the bolometric correction for $z=0$ as $k_{\textrm{bol,o}}=k_{\textrm{bol}}/18.7$. 

\subsubsection{Long GRBs}

The methods used in this work are applicable to all GRBs regardless of classification as long or short. As a comparison test case we include the gold (48) and silver (18) samples of LGRBs from \citet{Ghirlanda2017} with estimates on the deceleration time. \citet{Ghirlanda2017} also provide the redshift and the isotropic gamma-ray energy $E_\gamma$. The silver sample has poorly constrained prompt emission properties (e.g., $E_{p,\textrm{obs}}$) as the \textit{Swift}/BAT energy band (15-150 keV) and low measured flux lead to difficulty in accurately constraining their spectra. In order to overcome this, \citet{Ghirlanda2017} adopted estimates for the isotropic equivalent gamma-ray energy $E_\gamma$ and $E_{p,\textrm{obs}}$ for these 18 bursts from \citet{Butler2007,Butler2010} and \citet{Sakamoto2011} (see \citet{Ghirlanda2017} for details). 

For these 66 bursts, the estimate of $t_p$ comes from optical afterglows or the \textit{Fermi}/LAT GeV lightcurve. \citet{Ghirlanda2017} chose to use optical afterglow peaks because the X-ray data have contamination from plateaus or early-steep declines for many long GRBs. We do not perform the same analysis using optical afterglows from SGRBs due to the low completeness of optical afterglow detection from SGRBs ($\sim40\%$ when accounting for observational constraints; \citet{Fong2015}). 

\subsection{SGRBs Classified as observationally hostless}
\label{sec: obs-hostless observations section}

There have been 8 SGRBs detected by \textit{Swift} BAT that have been referred to as observationally hostless in the literature. These SGRBs are part of a small sample of sub-arcsecond  localized events that lack bright galaxies at the localization position (and therefore have no strong host association).
In Table \ref{Table_Hostless_Bursts}, we present these observationally hostless SGRBs along with the apparent magnitude limits on a coincident host galaxy. In \S \ref{sec_hostlesslimits}, we present the limits on circumburst density as well as further discussion on the nature of these observationally hostless bursts (see also Appendix \ref{Appendix: obs hostless}).
\par
We include 5/8 (63\%) of these SGRBs in our sample of \textit{Swift}/XRT observed bursts. We exclude GRBs 080503 and 090515 from our sample because their XRT lightcurves show potential contamination from prompt emission. In the case of GRB 090305A, this burst was not detected by \textit{Swift}/XRT despite the relatively short slew time of 103.4 seconds. The burst, however, did have an optical afterglow detected within the BAT error circle, which allowed for better localization and led to an XRT detection \citep{Cenko2009,Beardmore2009,Nicuesa2012,Tunnicliffe2014}. We exclude this burst from our X-ray data sample, as a single XRT detection does not fit our criteria. GRBs 080503 and 090515 were also detected in the optical. We discuss these optical detections in the larger picture of the circumburst density limits in \S \ref{sec_hostlesslimits}.

\subsection{SGRB XRT completeness}
\label{sec: completeness}
The intensity of a GRB afterglow is correlated with its environment \citep{Sari1998} with GRBs in low density environments leading to significantly fainter afterglows compared to those with higher circumburst densities \citep{Panaitescu2001,Salvaterra2010}. If we assume GRBs with only a single afterglow detection are from forward shock emission (specifically, synchrotron radiation in the slow cooling regime) then it is possible their faintness is due to a lower circumburst density (compared to the rest of the sample). In order to explore this possibility and justify excluding these GRBs from our sample, we investigate the XRT afterglow completeness of the \textit{Swift} SGRB sample as a function of their prompt $\gamma$-ray fluence (which is expected to be fully independent of the circumburst density).

\begin{figure} 
\centering
\includegraphics[width=\columnwidth]{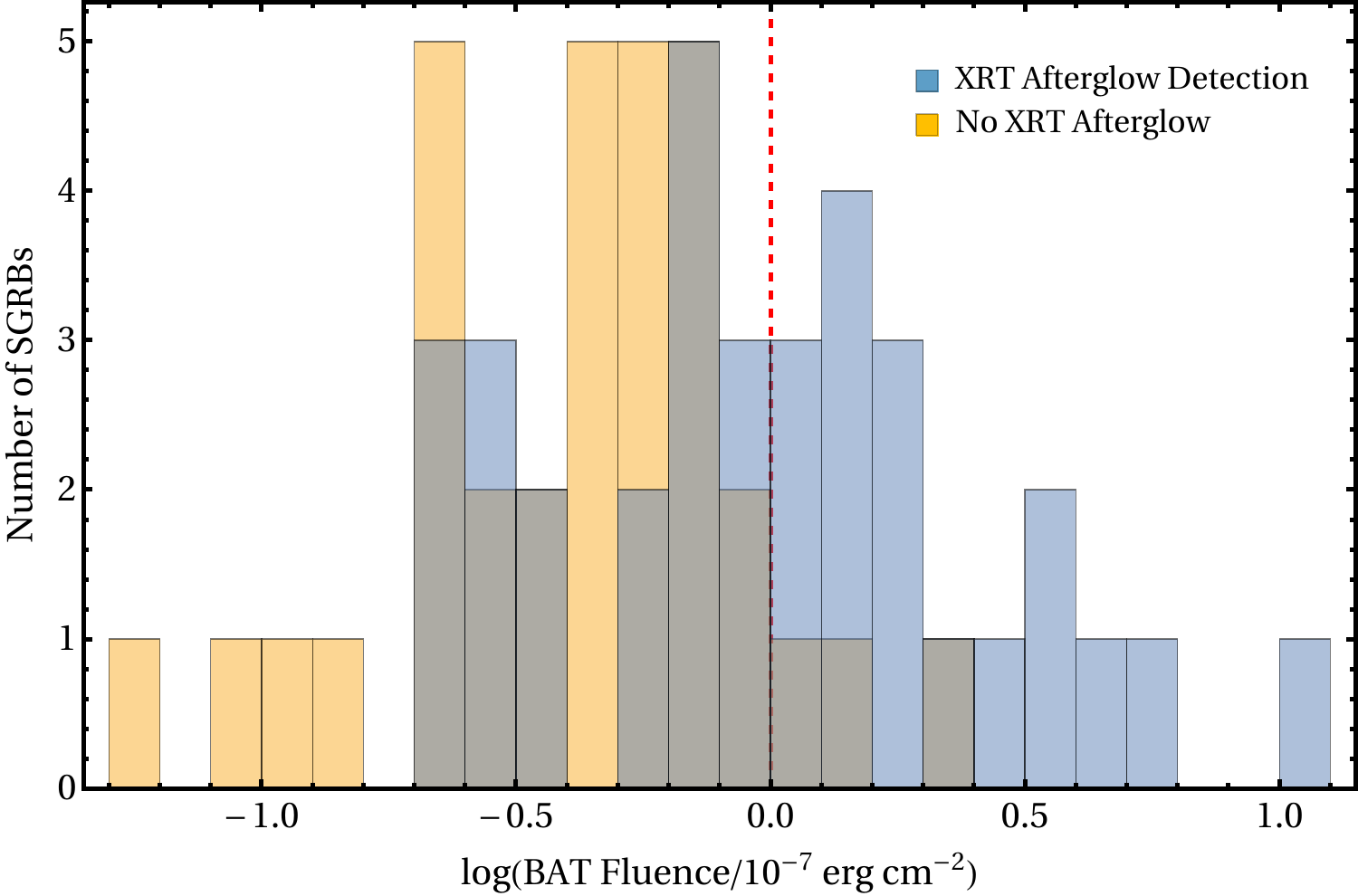}
\includegraphics[width=\columnwidth]{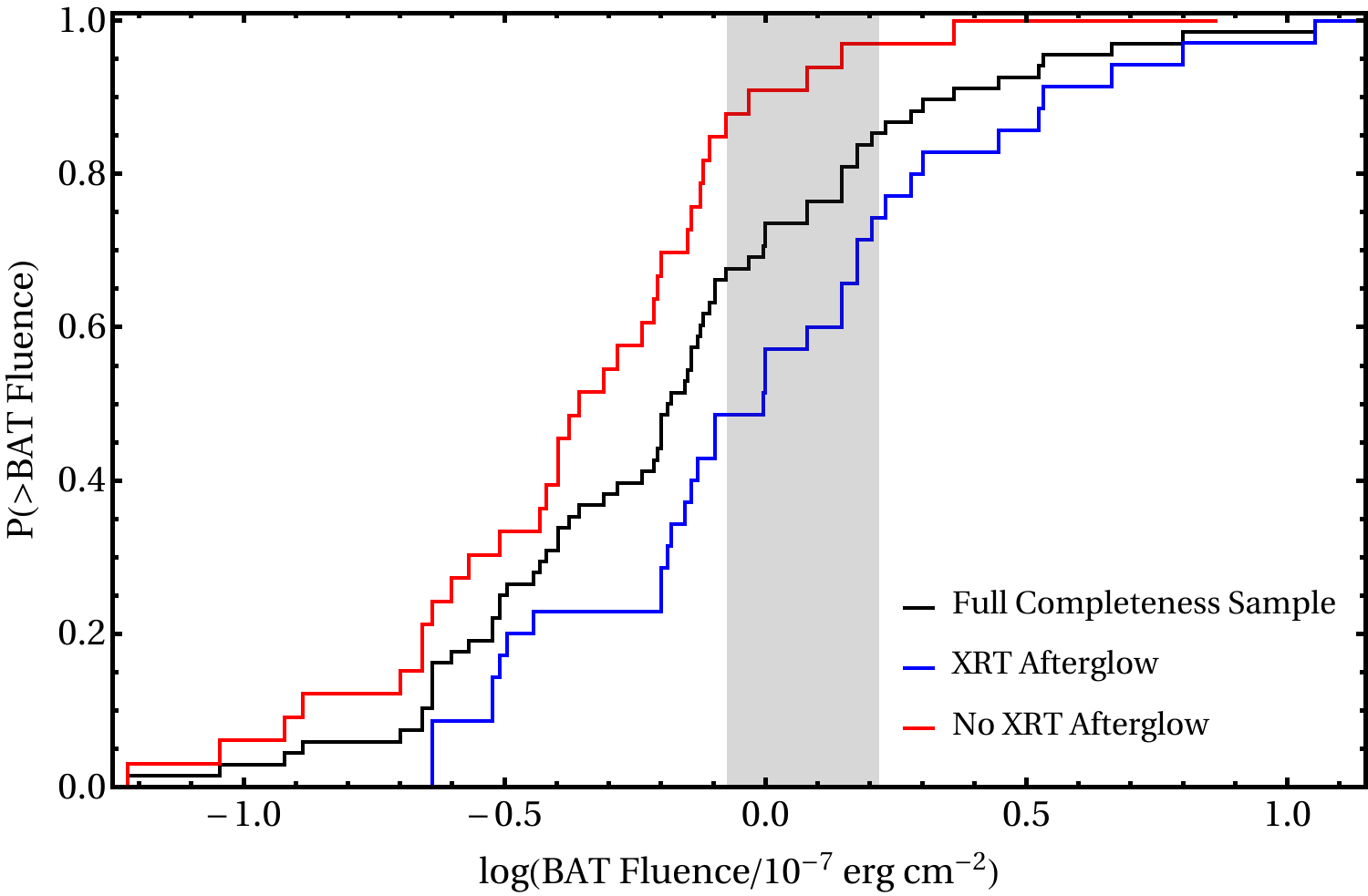}
\caption{(Top) Distribution of \textit{Swift} SGRBs without (yellow) XRT afterglow detection, as defined in this work, and with XRT afterglow detection (blue) versus their BAT fluence (15-150 keV). The dashed vertical line marks the median 90\% completeness fluence, $\phi_{\gamma,90}$, of XRT afterglow detection for \textit{Swift} SGRBs with $T_{90}\lesssim 0.8\, s$. (Bottom) Cumulative distribution function of BAT fluences for SGRBs in our completeness sample, both with and without XRT detection. The gray shaded region marks the $1\sigma$ confidence for the 90\% completeness fluence.} 
\label{SGRB_Completeness}
\end{figure}

There have been 118 (9\% of all \textit{Swift} GRBs) SGRBs detected by \textit{Swift} BAT with $T_{90}\lesssim 2\, \textrm{s}$, and only 88 (6.6\%) with $T_{90}\lesssim 0.8\, \textrm{s}$. Out of the 88 SGRBs, 8 (9.1\%) have not been observed by XRT (due to solar observing constraints) and 25 (27\%) have upper limits on an XRT detection. Of these 25, 12 were observed with XRT in less than 1000 seconds after the BAT trigger; we include these 12 events in our completeness sample of SGRBs\footnote{We justify this based on the fact that for SGRBs with XRT afterglow detections, the majority are still detected out to 1000 seconds.}. The remaining 13 SGRBs (which were observed with XRT after much longer intervals), and the 8 events lacking an XRT observation, were excluded from this analysis. This leaves us with a sample of 67 SGRBs of which the bursts in Table \ref{Table_SGRB_DATA} are a subset. We then determine the fraction for which the X-ray lightcurve shows a typical afterglow decay $\propto t^{-1}$ and find that only 34 (51\%) fulfill this criterion\footnote{We note that this analysis is limited to $T_{90}\lesssim 0.8\, \textrm{s}$, whereas the 52 SGRBs in Table \ref{Table_SGRB_DATA} also include those with $0.8 \lesssim T_{90}\lesssim 2\, \textrm{s}$.}, leaving 33 (49\%) events out of our final sample. Fig. \ref{SGRB_Completeness} (Bottom) exhibits the cumulative distribution of BAT fluence for SGRBs with XRT afterglow detections. 

Using our completeness sample, we determine the detectability of the X-ray afterglow for SGRBs as a function of their BAT fluence.  We find that the X-ray afterglow detectability rises rapidly as the BAT fluence increases, as is expected from the standard fireball model.
The completeness of XRT detectability is defined as the detected fraction of events above a given fluence, $f_\textrm{det}(>\phi_\gamma)=N_\textrm{det}(>\phi_\gamma)/N_\textrm{tot}(>\phi_\gamma)$. We apply a bootstrap algorithm to determine the fluence above which 90\% of SGRBs have XRT afterglow detections.  We obtain a 90\% completeness fluence for afterglow detections by XRT of $\phi_{\gamma,90}=1.2^{+0.5}_{-0.3}\times 10^{-7}\, \textrm{erg/cm}^2$ (see Fig. \ref{SGRB_Completeness}).

Fig. \ref{SGRB_Completeness} (Top) presents the SGRB distribution per BAT fluence. GRBs without afterglow detections (or with a single detection) have lower fluences compared to the rest of the SGRB population, although there is some overlap between the two distributions. We consider the former as evidence that their faint afterglows are due to lower overall burst energies or larger distances and not dominated by their environments. We, therefore, conclude that our sample is not significantly biased towards SGRBs residing in the highest density regions.

\subsection{11-Hour X-ray Flux}

The energy associated with prompt $\gamma-$rays has been shown to correlate linearly with the energy in the X-ray afterglow at 11-hours post-trigger \citep{Kumar2000,Freedman2001,Nysewander2008, Gehrels2008, Berger2013,Beniamini2016corr}. Assuming that electrons radiating in the X-ray band at 11-hours are in the slow cooling regime, then $F_{X,11}\propto n^{1/2}$ and the correlation can be used to constrain the width of the SGRB circumburst density distribution (see \S \ref{sec: 11-hour flux methods}). We compile the 11-hour X-ray flux, $F_{\textrm{X},11}$, in the $0.3-10$ keV band and the BAT fluence, $\phi_\gamma$, in $15-150$ keV band from the samples of \citet{Nysewander2008} and \citet{Berger2013}. \citet{Berger2013} does not provide the error for the 11-hour X-ray flux so we assume an error that is the average error from \citet{Nysewander2008}. There is significant overlap between those two samples, and due to the lack of error on 11-hour X-ray flux from \citet{Berger2013} we have chosen to include the values of \citet{Nysewander2008} for common bursts. We were able to add 8 SGRBs (see Table \ref{Table_11-hour_flux_bursts}) to this sample by extrapolating the X-ray lightcurves on the \textit{Swift}/XRT GRB Lightcurve Repository to 11-hours. In total, we include 22 bursts from \citet{Nysewander2008}, 14 from \citet{Berger2013}, and here-in added 8 bursts. We choose bursts within our sample (Table \ref{Table_SGRB_DATA}) that have X-ray detections at $t>10^{4}$ s allowing for extrapolation of the XRT lightcurve to 11-hours. We correct the absorbed fluxes using the PIMMS software, as above. 

\begin{table}
\centering
\caption{Estimates of X-ray flux at 11-hours post-trigger ($0.3-10$ keV) for 8 SGRBs, compiled by extrapolating the X-ray lightcurves on the \textit{Swift}/XRT GRB Lightcurve Repository.}
\begin{tabular}{|c|c|}
\hline
\hline
GRB &  $F_{X,11}$ ($10^{-13}$ erg/cm$^2$/s)  \\
\hline
\hline
130912A & $1.49_{-0.67}^{+1.23}$  \\[0.5mm]
131004A &$2.36_{-0.27}^{+0.31}$ \\[0.5mm]
140903A &$17.09_{-2.10}^{+2.40}$\\[0.5mm]
140930B &$1.16_{-0.34}^{+0.47}$ \\[0.5mm]
160525B & $1.01_{-0.18}^{+0.23}$\\[0.5mm]
160601A & $1.05_{-0.36}^{0.55}$\\[0.5mm]
160821B & $1.26_{-0.48}^{+0.77}$\\[0.5mm] 
160927A & $1.16_{-0.20}^{+0.24}$\\[0.5mm]
\hline
\end{tabular}
\label{Table_11-hour_flux_bursts}
\end{table}

\section{Method}
\label{sec: methods}

\subsection{Limits from Deceleration Peaks}
\label{sec: lower limits methods sec}
We apply the standard model of GRB afterglow emission \citep{Meszaros1997,Sari1998,Wijers1999} to X-ray afterglow detections in order to set lower limits on the circumburst density for our sample of SGRBs. In this model, the afterglow is due to synchrotron radiation from electrons accelerated in the forward shock, where the dynamics of the shock-wave are governed by the Blandford-McKee self-similar solution \citep{Blandford1976}. The shock-wave drives into an ambient medium density of the form $\rho_{\textrm{ext}}(r)=A\, r^{-k}$, where we use in what follows $k=0$ corresponding to a uniform density medium. In the case of LGRBs, with a massive star progenitor, a wind medium with $k=2$ is often more appropriate \citep{Granot2002,Schulze2011}. For SGRBs, it is not expected that a wind medium is applicable \citep{Sari1999,Granot2002,Soderberg2006}. 

\par
For each SGRB in our sample, we have acquired a lower limit, $F_{X,o}$, on the peak X-ray flux ($F_{X,o}<F_{X,p}$) and an upper limit, $t_o$, on the time of the afterglow peak ($t_p<t_o$)\footnote{In what follows, we consider all SGRBs to be viewed within the core of the jet (i.e., on-axis, $\theta_v<\theta_c$). This is a conservative assumption as the effect of increasing the viewing angle $\theta_v$ compared to the jet opening angle $\theta_c$ leads to lower Lorentz factors for material along the line of sight to the observer, which produces a later deceleration peak with a lower peak flux. This would lead us to infer larger densities. We note that even for a marginal off-axis viewing angle (e.g. $\theta_v/\theta_c=1.1$), we would observe shallow phases at early times as opposed to rapid increases \citep{Beniamini2019plateau}. } (see \S \ref{subsec: SGRBs Sample} and Table \ref{Table_SGRB_DATA}). The peak of the afterglow light-curve is also the time of deceleration.
The circumburst density can be established from the kinetic energy of the blastwave $E_{\textrm{kin}}$, initial bulk Lorentz factor of the jet $\Gamma$, and the deceleration time $t_p$ assuming an adiabatic flow \citep{Sari1999}. An expression for the initial bulk Lorentz factor \citep{Sari1999,Nappo2014,Ghirlanda2017} can be inverted for the circumburst density. Using $t_p<t_o$ we obtain a lower limit on the circumburst density
\begin{align}
n_{t_o}=\frac{17}{64\pi}\frac{E_\textrm{kin}}{c^5 \Gamma^8 m_p (\frac{t_o}{1+z})^3}\,\textrm{cm}^{-3},
\label{decellimprior}
\end{align}
where $c$ is the speed of light, and $m_p$ is the mass of a proton. We find the kinetic energy by converting from the isotropic equivalent $\gamma$-ray energy released in the prompt emission $E_\gamma$, 
\begin{align}
E_\textrm{kin}=E_\gamma \frac{1-\varepsilon_\gamma}{\varepsilon_\gamma}=E_\gamma\; \xi_\gamma,
\end{align}
where $\varepsilon_\gamma$ is the amount of energy in prompt gamma-rays relative to the total blastwave's energy. We have defined $\xi_\gamma=(1\!-\!\varepsilon_\gamma)/\varepsilon_\gamma$. We use a gamma-ray efficiency of $\varepsilon_\gamma=0.15$ \citep{Nava2014,Beniamini2015,Beniamini2016corr}. 
The isotropic equivalent $\gamma$-ray energy is
\begin{align}
E_\gamma=k_{\textrm{bol}} \frac{4\pi d_L^2}{1+z}\phi_\gamma \: \textrm{erg},
\label{eqn: Egamma }
\end{align}
where $d_L$ is the luminosity distance and $\phi_\gamma$ is the 15-150 keV BAT fluence. Combining equations (\ref{decellimprior}-\ref{eqn: Egamma }) yields a limit depending only on observables and on $\Gamma$
\begin{align}
n_{t_o}=\,2.8\,\frac{\phi_{\gamma,-7}d_{L,28}^2(1+z)^2 k_{\textrm{bol,o}}\xi_{\gamma,o}}{\Gamma_2^8\,  t^3_{o,2}}\,\textrm{cm}^{-3},
\label{decellim}
\end{align}
where we have used $\xi_{\gamma,o}=\xi_\gamma/5.7$. Here and elsewhere we adopt the convention $Q_x= Q/(10^x\,\textrm{cgs})$. We perform this calculation for the 19 SGRBs in our sample with redshift measurements.

\begin{figure} 
\centering
\includegraphics[width=\columnwidth]{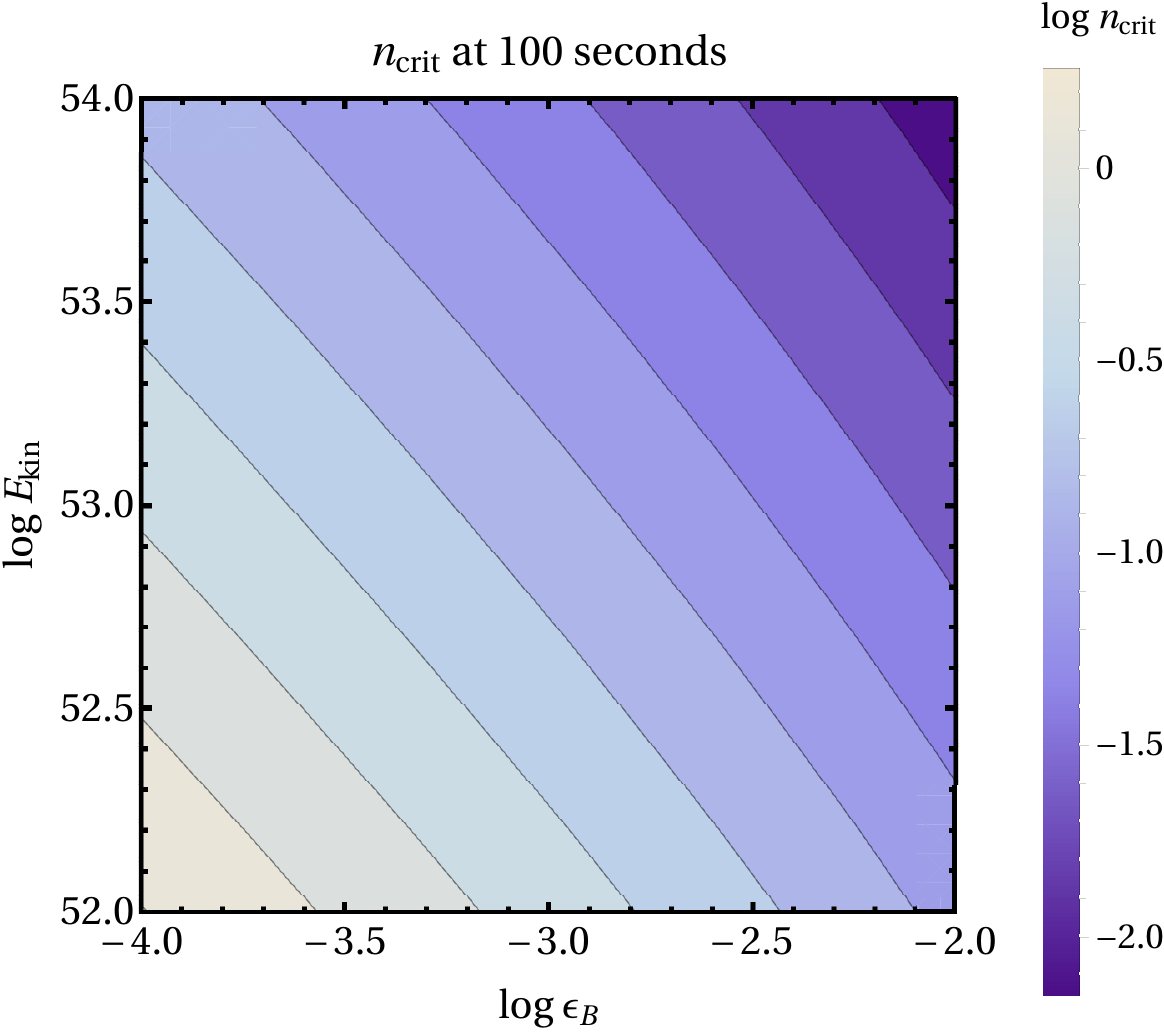}
\includegraphics[width=\columnwidth]{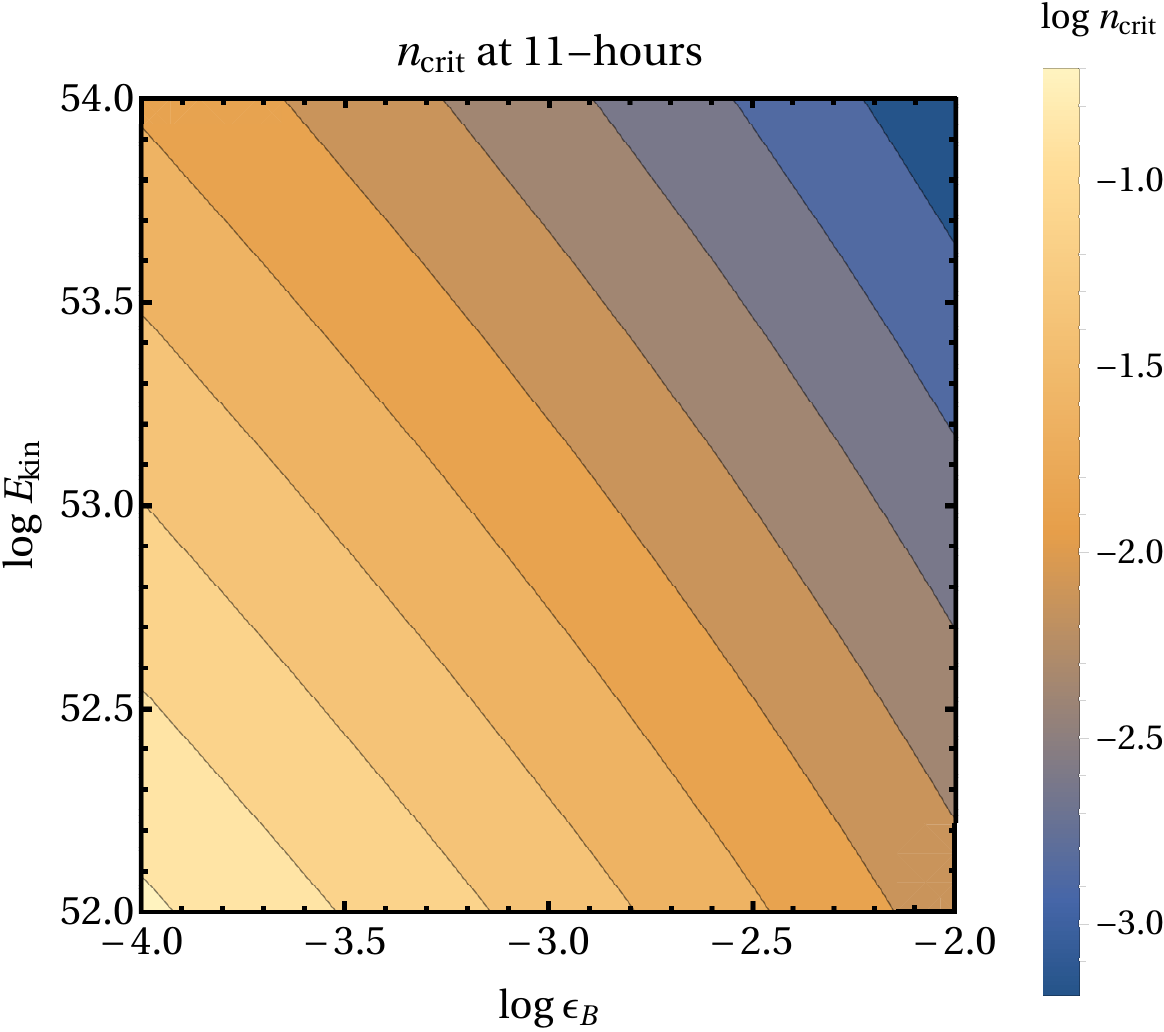}
\caption{Critical density at which the cooling frequency $\nu_c$ is equal to $1$ keV at (Top) 100 seconds and (Bottom) 11-hours post burst. For values of $n<n_{\textrm{crit}}$ a burst is in the slow cooling regime ($\nu_m<\nu<\nu_c$). These calculations were performed at $z=0$, and we note that the effect of increasing the redshift is to slightly decrease $n_\textrm{crit}\propto(1+z)^{-1/2}$ with an additional small effect from SSC corrections. For example, at $z=2$, $n_\textrm{crit}$ decreases by a factor of 2. 
}
\label{fig: crit density}
\end{figure}

\par
For electrons radiating synchrotron at the forward external shock in a uniform external density, $n$, the synchrotron cooling frequency (the frequency of radiation by electrons that cool on the dynamical time-scale of the system) is roughly anti-correlated with density $\nu_c\propto n^{-1}$ (there is a slight correction due to synchrotron self Compton, SSC; see below). Therefore, for bursts at low densities ($\sim \! 10^{-4}$ cm$^{-3}$) electrons radiating at $\nu\sim1$ keV are expected to be in the slow cooling regime $\nu_m<\nu<\nu_c$ at typical SGRB observation times (i.e., 100 seconds or 11-hours, see Fig. \ref{fig: crit density} below). Using the formulation of \citet{Granot2002}, and including Inverse Compton (IC) corrections \citep{Sari2001,Zou2009,Beniamini2015}, the cooling frequency $\nu_c$ and the injection frequency of electrons $\nu_m$ are given by
\begin{align}
    &\nu_c=8.6\times 10^{22}\, \varepsilon_{B,-4}^{-3/2}n^{-1}_{-4}E_{\textrm{kin},52}^{-1/2}t_{\textrm{days}}^{-1/2}(1+z)^{-1/2}(1+Y)^{-2}\; \textrm{Hz}, \label{eqn: slow cool frequ.}\\
    &\nu_m=1.6\times 10^{10}\,E_{\textrm{kin},52}^{1/2}\varepsilon_{B,-4}^{1/2}\varepsilon_{e,-1}^{2}t_{\textrm{days}}^{-3/2}(1+z)^{1/2}\; \textrm{Hz},
\end{align}
where $\varepsilon_B$ and $\varepsilon_e$ are the fractions of the burst kinetic energy during the afterglow phase that reside in the magnetic field and electrons respectively. The IC parameter is given by \citet{Beniamini2015}:
\begin{align}
\label{eqn: Yparam globally slow cooling}
    Y=10\,\varepsilon_{e,-1}^{2/3}\varepsilon_{B,-4}^{-4/9}\Big(\frac{E_{\textrm{kin},52}\,n_{-4}\,(1+z)}{t_{\textrm{days}}}\Big)^{1/20},
\end{align}
where we have used a typical value of $p=2.2$ for the slope of the electrons' initial power-law energy distribution $N(\gamma)\propto \gamma^{-p}$ for $\gamma\gtrsim\gamma_m$ with $\gamma_m$ corresponding to the energy of electrons that radiate at $\nu_m$. In order to ensure the assumption of slow cooling is consistent, we calculate a critical density $n_{\textrm{crit}}$ at which the cooling frequency is equal to $\nu_\textrm{keV}=\nu/\textrm{keV}$:
\begin{align}
    n_\textrm{crit} = 0.17\, \Big(\varepsilon_{B,-4}^{5/9}\, \varepsilon_{e,-1}^{40/33}t_\textrm{days}^{4/11}\,\nu_\textrm{keV}^{10/11}\, E_{\textrm{kin},52}^{6/11}\, (1+z)^{6/11}\Big)^{-1}\,\textrm{cm}^{-3}.
\label{eqn: crit density}
\end{align}
For values of density less than this critical density  ($n<n_{\textrm{crit}}$) the electrons radiating at $\nu_\textrm{keV}$ are in the slow cooling regime. In Fig. \ref{fig: crit density}, we show the critical density at both 100 seconds and 11-hours post burst for typical kinetic energies and values of $\varepsilon_B$. At 100 seconds post burst we find that for the low density environments considered in this work, practically all SGRBs reside in the slow cooling regime. We also find that at 11-hours the assumption of slow cooling is typically correct for standard expected values of the SGRB parameters. This becomes relevant to our discussion in \S \ref{sec: 11-hour flux corr}.
\par

\par
Due to the fact that the cooling frequency is very large, the peak of $\nu L_\nu$ at $\nu_c$ will not be observable for bursts in low density environments. The observed emission will be dominated by the highest observable frequency, generally in the X-ray band. We integrate the spectral luminosity over the \textit{Swift} frequency range (0.3-10 keV) including the IC corrections. The total observed luminosity is 
\begin{align}
   L_X=& \int^{\nu_U = 10\, \textrm{keV}}_{\nu_L =0.3\, \textrm{keV}} \Big(\frac{\nu}{\nu_c} \Big)^{(1-p)/2}\, L_{\nu_c}d\nu, \nonumber \\
 =&\, 2.5\times 10^{42}\, \frac{\varepsilon_{e,-1}^{6/5}\varepsilon_{B,-4}^{4/5}\, E_{\textrm{kin},52}^{1.3}n_{-4}^{1/2}}{t_{\textrm{days}}^{0.9}}(1+z)^{1.3}\, \textrm{erg/s}, \label{eqn: Lx not peak yet}
\end{align}
where we have again used $p=2.2$. The luminosity peaks at the deceleration time $t_p$ which by rearranging equation (\ref{decellim}) is:
\begin{align}
\frac{t_p}{1+z}=0.72 \Big(\frac{E_{\textrm{kin},52}}{n_{-4}} \Big)^{1/3}\Gamma_2^{-8/3}\,\textrm{hours}.
\end{align}
Inserting this into the observed X-ray flux $L_X$ yields an expression for the peak afterglow X-ray luminosity $L_{X,p}$ of a GRB in the slow cooling regime ($\nu_m<\nu<\nu_c$):
\begin{align}
L_{X,p}=5.6\times 10^{43}\; \varepsilon_{e,-1}^{6/5}\varepsilon_{B,-4}^{4/5}\, E_{\textrm{kin},52} \,n_{-4}^{4/5}\Gamma_2^{12/5}(1+z)^{2/5}
\textrm{erg/s}.
\label{peaklumSGRB}
\end{align}
In the slow cooling regime, where $\nu_c\gg\nu_m$, electrons radiate synchrotron inefficiently. Coupled with the fact that $\nu_c\gg\nu\approx 1$ keV (implying that most of the emitted energy is missed by the observer), this predicts that the flux for low density events will be small compared to other SGRBs. Thus, a lower limit on the true peak flux of the afterglow allows us to set meaningful lower limits on the density.
\par
 Throughout our calculations we adopt a constant value of $\varepsilon_{e,-1}=1$, which corresponds to 10\% of the blastwave energy being initially held in relativistic electrons \citep{BeniaminiVanDerHorst}. The fraction of energy in the magnetic field $\varepsilon_B$ is a less well constrained quantity. We adopt a value of $\varepsilon_B=10^{-2}$ while noting that a more likely value is  $\varepsilon_B=10^{-4}$ \citep{Barniol2014,Santana2014,Zhang2015}. We choose the larger value because it leads to a more conservative lower limit on the circumburst density $n$ (by a factor of 100). We use the lower limit on the peak X-ray flux ($F_{X,p}=F_{X,o}$) and invert equation (\ref{peaklumSGRB}) to find 
\begin{align} 
n_{F_{X,o}}
=&\,1.1\, \Big(\frac{F_{X,o,-10}}{k_{\textrm{bol,o}}\,\xi_{\gamma,o\,\phi_{\gamma,-7}}}\Big)^{5/4}\frac{(1+z)^{3/4}}{\varepsilon_{e,-1}^{3/2}\varepsilon_{B,-4}\Gamma^3_2} \; \textrm{cm}^{-3}.
\label{peaklumlimit}
\end{align}
The big advantage of this method is that it depends only very weakly on the distance (through the $1+z$ term, but without an explicit dependence on $d_L$). Specifically, 
equation (\ref{peaklumlimit}) has a redshift dependence that is minimized at $z=0$. 
\par
The parameter with the largest uncertainty and most important effect on these lower limits is the bulk Lorentz factor. The lower limits on density have a very strong negative dependence on the bulk Lorentz factor: $n_{t_o}\propto \Gamma^{-8}$ and $n_{F_{X,o}}\propto \Gamma^{-3}$. In other words, smaller values of $\Gamma$ lead to stronger limits. Thus, it is important to choose a value of $\Gamma$ that is both in line with current research and conservative in terms of our limits. Much work has been done to constrain the bulk Lorentz factor of LGRBs using observations of the deceleration time in the X-rays and Optical \citep{Ghirlanda2012,Lu2012,Liang2010}. Using deceleration peaks observed in the optical for sample of LGRBs, \citet{Ghirlanda2017} found that the average for long bursts is $\Gamma=320$. Throughout this work we adopt a value $\Gamma=300$ in our calculations. We note that compactness arguments, as applied to SGRBs, yield more modest values for $\Gamma$ relative to those for LGRBs \citep{Zou2011,Hascoet2012,Nava2017}. 
Thus, adopting the average bulk Lorentz factor of a sample of LGRBs is expected to be an overestimation of the true values and thus conservative for the purposes of limiting the density. 
\par
For bursts with redshift measurements, we can combine our limits based on the peak time and flux in order to remove uncertainty in the bulk Lorentz factor by using equation (\ref{decellimprior})
\begin{align}
\Gamma>350\,\Big(\frac{E_{\textrm{kin},52}(1+z)^3}{n_{-4}\,t_{o,2}^3}\Big)^{1/8}
\end{align}
which we insert into equation (\ref{peaklumSGRB}) for the peak luminosity,
\begin{align}
    n>n_{\textrm{comb}} =&\,0.7\,\frac{t_{o,2}^{1.8}F_{X,o,-10}^2}{\varepsilon_{e,-1}^{12/5}\varepsilon^{8/5}_{B,-4}d_{L,28}^{1.2}(k_{\textrm{bol,o}}\xi_{\gamma,o}\phi_{\gamma,-7})^{2.6}}\; \textrm{cm}^{-3}.
    \label{combinedlimit}
\end{align}
We note that direct rearrangement of equation (\ref{eqn: Lx not peak yet}) yields the same result. Equation (\ref{combinedlimit}) sets a limit on the circumburst density that is independent of $\Gamma$ but does rely on choices of the fraction of energy in the electrons and magnetic field.

\par
In order to identify the most constraining lower limit on circumburst density for each burst, we must take into account equations (\ref{decellim}), (\ref{peaklumlimit}), and (\ref{combinedlimit}). The minimum permitted density from the limits on the time and flux at the afterglow peak in the slow cooling regime is 
\begin{align}
\label{eqn: nlim}
    n_\textrm{lim}=\max\big(n_{t_o},\, n_{F_{X,o}},\, n_\textrm{comb}\big).
\end{align}
We have shown that the majority of SGRBs will be in the slow cooling regime at the time of the afterglow peak (Fig. \ref{fig: crit density}). But, if a burst is not in the slow cooling regime at the time of these limits, $t_o$, then the most conservative approach we can take is to apply equation (\ref{eqn: crit density}) ($n_\textrm{crit}$) as the lower limit on density. This is because in the fast cooling regime, the flux is independent of density and we cannot obtain a  more constraining limit\footnote{We note that the limit from the peak time is not constrained by $n_\textrm{crit}$, as it is independent of whether electrons radiating in the X-rays are slow cooling. However, it is conservative for the purposes of lower values of the bulk Lorentz factor (e.g., $\Gamma\sim100$; where the peak time limit is most constraining) as both the combined limit and critical density are independent of the choice of $\Gamma$.} than $n_\textrm{crit}$. Therefore, to take into account this possibility the overall lower limit on density is 
\begin{align}
    n_\textrm{min}=\min\big(n_\textrm{lim},\, n_\textrm{crit}\big).
    \label{eqn: true nmin}
\end{align}
This limit, $n_{\rm min}(z)$, is minimized at some finite $z_\textrm{min}>0$, see Fig. \ref{obshostlesslimitsboth} for an example. The simplest, and most conservative approach is simply to take $n_{\rm min}(z_{\rm min})$ as the limit for SGRBs without redshift measurements. However, since there is no reason to expect that $z_{\rm min}$ is the true redshift of a given SGRB and due to the steep derivative of $|\frac{\partial n_\textrm{min}}{\partial z}|_{z_{\rm min}}$, this method is grossly over-conservative. We employ instead a statistical approach as detailed below. The latter is justified from a sample point of view (recall that we have 52 SGRBs in our overall sample) and is robust so long as statistical fluctuations are well accounted for. We calculate the expected value of $n_\textrm{min}$ and its standard deviation using redshift distributions $P(z)$ from WP15 and \citet{Ghirlanda2016} (hereafter, G16):
\begin{align}
   \langle n_\textrm{min}\rangle = \int^{10}_{0}P(z)\,n_\textrm{min}(z)\,dz \quad \, \quad \sigma_{n_{\rm min}}= \langle n^2_\textrm{min}\rangle- \langle n_\textrm{min}\rangle^2.
   \label{eqn: average nmin}
\end{align}
The redshift distributions from WP15 and G16 are in the source frame. In order to convert them to the observer frame, we use a convolution of the comoving rate per unit volume $\Psi(z)$ and the gamma-ray luminosity function $\phi(L)$: 
\begin{align}
   P(z) = \frac{\Psi(z)}{1+z} \frac{dV}{dz} \int_{L_\textrm{min}(z)} \phi(L) \,dL,
\end{align}
where $dV(z)/dz$ is the comoving volume element. The minimum detectable gamma-ray luminosity at a given redshift $L_\textrm{min}(z)$ is determined by the minimum detectable flux; we adopt $F_\textrm{min}=5\times 10^{-8}$ erg cm$^{-2}$ s$^{-1}$ for \textit{Swift}/BAT in the $15-150$ keV band \citep[see Fig. 8 in ][]{Lien2014}.

\subsection{Limits from 11-Hour Flux Correlation}
\label{sec: 11-hour flux methods}

A tight correlation between the X-ray afterglow luminosity at 11-hours and the $\gamma$-ray energy of GRBs (or equivalently between the X-ray flux at 11-hours and the $\gamma$-ray fluence) is to be expected, if the X-ray luminosity is dominated by electrons radiating in the fast cooling regime ($\nu>\nu_c,\nu_m$) and if SSC corrections to the cooling can be ignored (i.e., $Y\ll 1$) \citep{Kumar2000,Freedman2001,Nysewander2008, Gehrels2008, Berger2013,Beniamini2016corr}.
Accounting for SSC cooling, the X-ray flux in the fast cooling regime is 
\begin{align}
\label{eqn: 11-hour flux FC}
    F_{X,11}\!=\!1.8\times10^{-12}\phi_{\gamma,-7}^{1.05}\; \varepsilon_{e,-1}^{6/5} \varepsilon_{B,-4}^{1/20} d_{L,28}^{1/10} \frac{(k_{\textrm{bol,o}}\xi_{\gamma,o})^{1.05}}{1+Y}\textrm{erg/cm$^{2}$/s}
\end{align}
where, as above, we have adopted $p=2.2$. The Compton parameter for $\nu_m<\nu_c$ is given by equation (\ref{eqn: Yparam globally slow cooling}) and for $\nu_c<\nu_m$ becomes \citep{Sari1996}:
\begin{align}
\label{eqn: Yparam globally fast cooling}
    Y= \begin{cases} 
      \sqrt{\frac{\varepsilon_{e}}{\varepsilon_{B}}}, & \textrm{for}\: \varepsilon_e>\varepsilon_B  \\
   \frac{\varepsilon_{e}}{\varepsilon_{B}}, & \textrm{for}\varepsilon_e<\varepsilon_B.
   \end{cases}
\end{align}
In the slow cooling regime, $\nu_m<\nu<\nu_c$, a similar correlation, but with a slightly different slope, is also expected \begin{equation}
\label{eqn: 11-hour flux SC}
    F_{X,11}\!=\!3.8\times 10^{-15}  \phi_{\gamma,-7}^{1.3}\; n_{-4}^{1/2} \varepsilon_{e,-1}^{6/5} \varepsilon_{B,-4}^{4/5} d_{L,28}^{3/5}(k_{\textrm{bol,o}}\xi_{\gamma,o})^{1.3}\: \textrm{erg/cm$^{2}$/s}.
\end{equation}
The normalization and width of the correlations depend on both the jet micro-physics (e.g., $\varepsilon_B$) and the circumburst density. In the fast cooling regime, we find $F_{X,11}\propto \phi_\gamma^{1.05}\varepsilon_B^{1/20}(1+Y)^{-1}$. This correlation will manifest in observed GRBs, assuming that $\epsilon_B$ does not vary widely between bursts (discussed further in \S \ref{sec: 11-hour flux corr}). For slow cooling, $F_{X,11}\propto \phi_\gamma^{1.3}\varepsilon_B^{4/5}n^{1/2}$. Here as well, a correlation may be seen in observed GRBs. This time, the requirement is that $\epsilon_B$, $n$ and $d_L$ should all not vary significantly between bursts. In either case (fast or slow cooling) the external density is constrained. For the first, we must have $n>n_{\rm crit}$ (see equation \ref{eqn: crit density}), while for the latter, the width of the density distribution, $\sigma_{\log n}$, should be sufficiently low to reproduce a correlation.

\citet{Berger2013} presented a correlation between the X-ray luminosity at 11-hours post-burst and the isotropic equivalent gamma-ray energy for both LGRBs and SGRBs. \citet{Berger2013} identified that the ratio $L_{X,11}/E_\gamma$ for LGRBs and SGRBs is consistent with being sampled from the same underlying distribution, with a Kolmogorov-Smirnov  test p-value of $p=0.23$. Since for LGRBs the FC regime is expected to be more relevant \citep[see, e.g.,][]{Beniamini2016corr}, 
the observed correlations in $L_{X,11}-E_\gamma$ (and $F_{X,11}-\phi_\gamma$) for LGRBs and SGRBs set a requirement on the typical circumburst density. This method has been previously used by \citet{Nysewander2008} to compare the ratio $F_{\textrm{opt}}/F_X$ at 11-hours for long and short bursts, where $F_\textrm{opt}$ is the burst optical flux. \citet{Nysewander2008} determined that in order to produce the observed ratio $F_\textrm{opt}/F_X$, the circumburst densities of SGRBs must be similar to the typical values for LGRBs of $\sim 1$ cm$^{-3}$. We will use the observed $F_{X,11}-\phi_\gamma$ correlation to set an additional limit on the density distribution of SGRBs (as well as the width of the distribution of $\varepsilon_B$).  In \S \ref{sec: 11-hour flux corr} we discuss the observed correlation and the limits it provides on the distribution of circumburst densities for SGRBs.

\subsection{Gas Density Profile}
\label{sec: gas density profile}

We have described methods to constrain the merger environments of SGRBs using lower limits on circumburst density. For SGRBs, as mentioned in \S \ref{sec: lower limits methods sec}, the external density probed by the GRB blast wave is most likely dominated by the local interstellar medium (ISM). Therefore, the inferred circumburst density of SGRBs traces the ISM environment at their merger locations and in turn the physical offset from the true host galaxy's center. The intrinsic offset distribution is uncertain due to the relatively poor (compared to optical) localization of SGRBs with \textit{Swift}/XRT. \citet{Cobb2008} identified, for a sample of 72 LGRBs localized by XRT, that $\sim 50\%$ of the galaxy's coincident with localization regions are likely false associations with random galaxies. In contrast, they found there is typically a $\sim 1\%$ chance of a random galaxy overlapping with an optical afterglow localization region. However, even for SGRBs with accurate localizations, the true offset is potentially overestimated due to the chance for missed galaxies (e.g., less massive or at higher redshift) at smaller offsets, which is explored in future work. In order to constrain the intrinsic offset distribution, we convert our lower limits on circumburst density to maximal physical offsets. We do this by assuming a radial density profile for ISM gas in a typical SGRB host galaxy.

We apply a gas density profile that is a combination of two components. The first component follows the observed surface density profile of neutral (HI + H$_2$) gas within galaxies, which closely follows the stellar surface brightness distribution. \citet{Bigiel2012} identified a universal neutral gas density profile using observations of galaxies from The HI Nearby Galaxy Survey (THINGS) and the HERA CO-Line Extragalactic Survey (HERACLES). \citet{Kravtsov2013} performed a similar analysis and determined that the gas distribution is $\sim 2.6$ times as extended as the stellar distribution. We apply the universal gas density profile derived by \citet{Kravtsov2013} where an exponential profile was assumed. We note that \citet{Kravtsov2013} identified that the mean stellar profile of early- and late-type galaxies is quite similar outside the half-mass radius\footnote{The half-mass radius can be used as a proxy for the 3D half-light radius $r_e$ with  $\sim25\%$ error \citep{Szomoru2013}.}.
This component of the density profile is normalized using the gas fraction ($M_g/M_*$) relation from \citet{Peeples2014}.

The second component is the MB gas model \citep{MB2004} which describes a hot gas halo under the assumption that the gas is adiabatic ($\gamma=5/3$) and is in hydrostatic equilibrium with the dark matter (i.e., NFW profile \citet{NFW1997}). The MB profile is given by \citep{MB2004,Fang2013}:
\begin{align}
    \rho_g^{\textrm{MB}}(x)=\rho_\textrm{vir}\Big[1+\frac{3.7}{x}\ln(1+x)-\frac{3.7}{c_\textrm{vir}}\ln(1+c_\textrm{vir})\Big]^{3/2},
\end{align}
where $x=r/r_s$ and $\rho_\textrm{vir}$ is the gas density at $r_\textrm{vir}$. The virial radius $r_\textrm{vir}$ is defined by the overdensity $\Delta_\textrm{vir}$ required for collapse of a spherical top-hat \citep{BN1998}, and the half-light radius is calculated using the $r_e-r_{200}$ relation from \citet{Kravtsov2013}. Conversion between stellar mass and virial mass, $M_\textrm{vir}$, is performed using the stellar mass-halo mass relation (SHM) from \citet{Moster2013}, where we also convert $M_{200}$, defined by $\Delta_{200}=200$, to $M_\textrm{vir}$ using the method of \citet{Hu2003}. The NFW profile \citep{NFW1997} scale radius $r_s$ is determined by the concentration parameter $c_\textrm{vir}$ \citep{Bullock2001}. Observations \citep{Humphrey2006,Humphrey2011,Humphrey2012,Buote2016} and simulations \citep{Barnes2017} of isolated elliptical galaxies, galaxy clusters, and fossil groups have shown that the gas mass fraction approaches the Universal baryon fraction at the virial radius, which demonstrates baryonic closure of virialized systems. Thus, we normalize the total gas density profile using the Universal baryon fraction: $M_g(r_\textrm{vir})=M_\textrm{vir}/6$. This normalization determines the density at the virial radius $\rho_\textrm{vir}$ in the MB model. 

Using this profile, we calculate the distribution of gas densities at $r_e$, $5r_e$, and $r_\textrm{vir}$, assuming the observed stellar mass distributions of SGRBs \citep{Leibler2010,FongBerger2013}, in order to constrain the merger location of SGRBs using the lower limits on density. We adopt a value of $n_\textrm{vir}=10^{-4}$ cm$^{-3}$ as the gas density at the virial radius. This value is comparable to the density at the Milky Way's virial radius \citep{Fang2013}. At the radii $r_e$ and $5 r_e$ we find typical densities of $10^{-1}$ cm$^{-3}$ and $10^{-2}$ cm$^{-3}$ respectively (assuming the median observed SGRB host galaxy mass). We emphasize that while these densities are highly model dependent\footnote{The density profile employed here should be taken as an average profile, while noting that the local ISM density may show significant fluctuations in density. However, simulations of ISM turbulent mixing driven by supernova explosions \citep{Korpi1999,Gent2013} have shown $10^{-4}$ cm$^{-3}$ is a reasonable lower bound on ISM density.}, the lower limits on SGRB circumburst density are model independent constraints on the radial gas density profile of SGRB host galaxies.

\section{Results}
\label{sec_Results}

\begin{table}
\centering
\caption{Density constraints considered in this paper.}
\begin{tabular}{|c|c|c|}
\hline
\hline
Notation & Definition & Equation   \\
\hline
$n_{t_o}$ & Lower limit from time of first detection & \ref{decellim} \\
$n_\textrm{crit}$ &  Maximum density for slow cooling regime & \ref{eqn: crit density} \\
$n_{F_{X,o}}$ & Lower limit from X-ray flux of first detection & \ref{peaklumlimit} \\
$n_\textrm{comb}$ & Combined lower limit from time and flux & \ref{combinedlimit}\\
$n_\textrm{min}$ & Overall lower limit (accounting for $n_\textrm{crit}$) & \ref{eqn: true nmin} \\
 $\langle n_\textrm{min}\rangle$ & Average limit integrated over redshift & \ref{eqn: average nmin}\\
\hline
\end{tabular}
\label{tab: Table_Equation_Summary}
\end{table}

\begin{table}
\centering
\caption{Upper limits on the fraction of SGRBs, $f(<n)$, consistent with circumburst density, $n$, below a given threshold  for all bursts (52) and those with redshift (19). 
For bursts without redshift we apply the redshift $z_\textrm{min}$ which minimizes the overall limit (equation (\ref{eqn: true nmin})). In equation (\ref{eqn: average nmin}), we average over the redshift distributions from \citet{Wanderman2015} (WP15) and \citet{Ghirlanda2016} (G16) respectively.
In these calculations we adopt $\varepsilon_e=0.1$, $\varepsilon_\gamma=0.15$, and $\Gamma=300$. 
}
\begin{tabular}{|c|c|c|c|c|}
\hline
\hline
 Equation & Sample & $\varepsilon_B$ & $f(<10^{-6})$ &$f(<10^{-4})$ \\
 (\ref{eqn: true nmin}): $n_\textrm{min}(z)$    &  with redshift &  $10^{-2}$  & 0  & 0.16 \\
  (\ref{eqn: true nmin}): $n_\textrm{min}(z)$  &  with redshift &  $10^{-4}$  & 0 & 0 \\
  (\ref{eqn: true nmin}): $n_\textrm{min}(z_\textrm{min})$  &  all bursts &  $10^{-2}$  & 0.02 & 0.32 \\
  (\ref{eqn: true nmin}): $n_\textrm{min}(z_\textrm{min})$ &  all bursts &  $10^{-4}$  & 0 & 0 \\
   (\ref{eqn: average nmin}): $\langle n_\textrm{min}\rangle$, WP15 &  all bursts &  $10^{-2}$  & 0 & 0.17\\
   (\ref{eqn: average nmin}): $\langle n_\textrm{min}\rangle$, WP15  &  all bursts &  $10^{-4}$  & 0 & 0 \\
      (\ref{eqn: average nmin}): $\langle n_\textrm{min}\rangle$, G16 &  all bursts &  $10^{-2}$  & 0 & 0.096 \\
   (\ref{eqn: average nmin}): $\langle n_\textrm{min}\rangle$, G16  &  all bursts &  $10^{-4}$  & 0 & 0\\
\hline
\end{tabular}
\label{Table_Full_Results}
\end{table}

\subsection{Limits on Circumburst Density}
\label{sec: density results}
\subsubsection{Sample with Redshifts}
\label{sec: density results 20 sample}

\begin{figure} 
\centering
\includegraphics[width=\columnwidth]{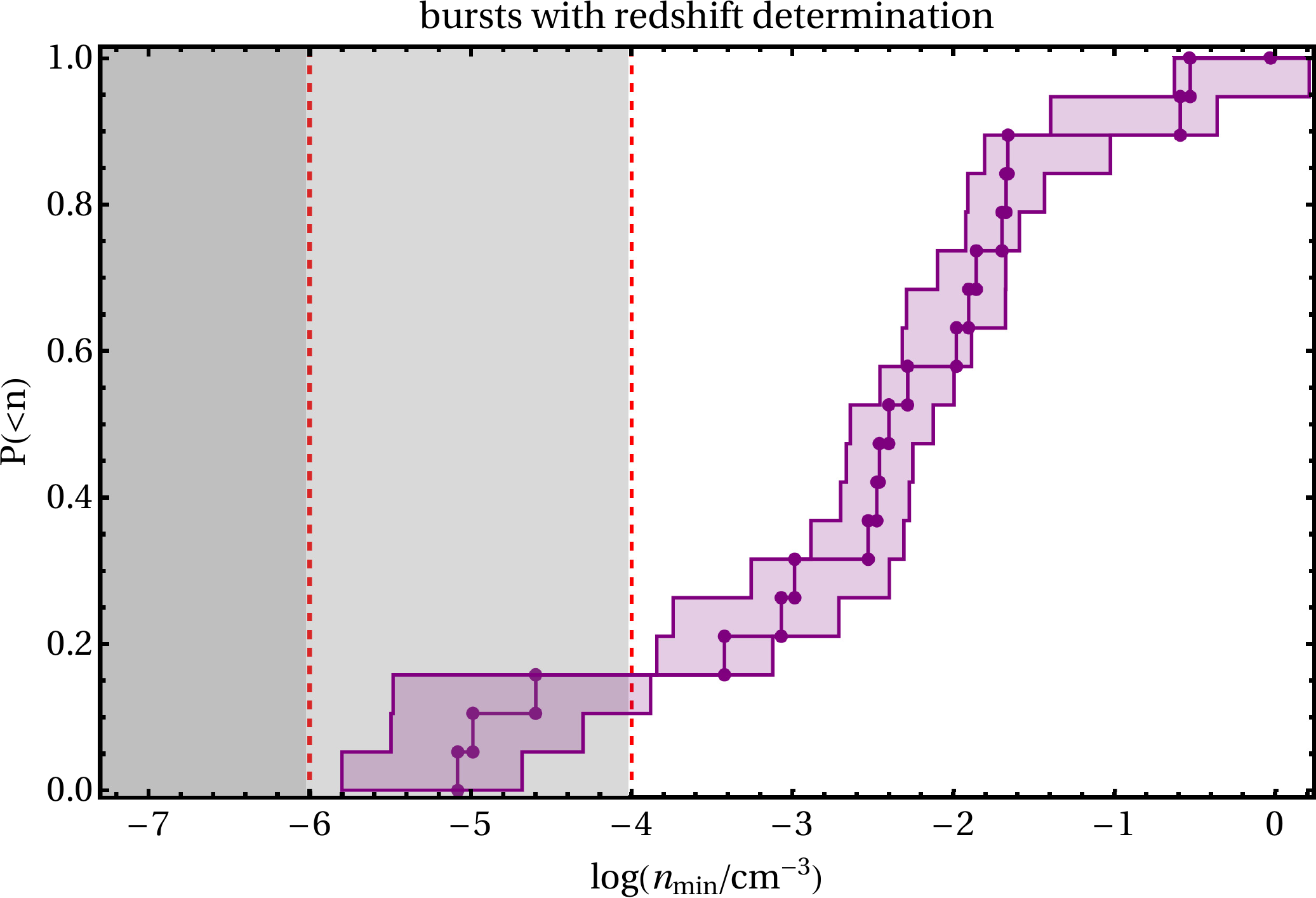}\label{fig:best limit}
\includegraphics[width=\columnwidth]{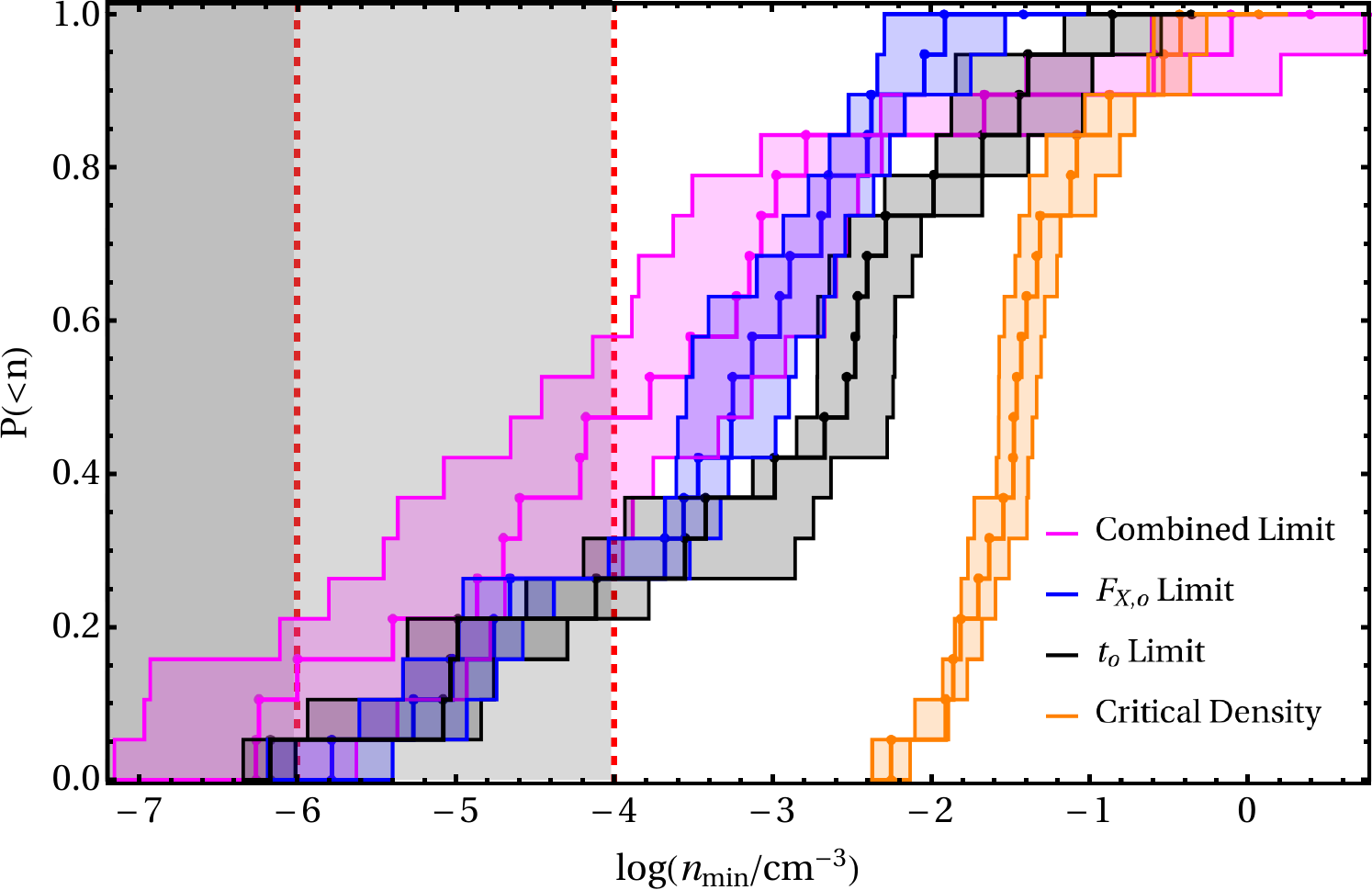}\label{fig:3CDFs}
\caption{Cumulative distribution functions for the lower limits on circumburst density calculated using $\Gamma=300$, $\varepsilon_B=10^{-2}$ for bursts with redshift determination. The dotted vertical lines, and shaded regions, represent densities consistent with $n_\textrm{min}\lesssim 10^{-6}\, \textrm{cm}^{-3}$ and $n_\textrm{min}\lesssim 10^{-4}\, \textrm{cm}^{-3}$, respectively. 
(Top) The most constraining lower limit (equation (\ref{eqn: true nmin})) for each of the 19 SGRBs with redshift is shown. (Bottom) Cumulative distribution functions for the deceleration time limit (equation (\ref{decellim})), peak luminosity limit (equation (\ref{peaklumlimit})), combined limit (equation (\ref{combinedlimit})), and critical density (equation (\ref{eqn: crit density})).  We emphasize that both equations (\ref{eqn: crit density}) and (\ref{combinedlimit}) are independent of $\Gamma$.
}
\label{hi}
\end{figure}

We present lower limits on circumburst density for our sample of 19 SGRBs with redshift (Table \ref{Table_SGRB_DATA}). In Fig. \ref{hi} (bottom), we demonstrate lower limits calculated using $n_{t_o}$, $n_{F_{X,o}}$, and $n_\textrm{comb}$. We present a summary of the relevant limits on density in Table \ref{tab: Table_Equation_Summary}. We stress that the limit from $n_\textrm{comb}$ is independent of bulk Lorentz factor, which is the largest source of uncertainty. The overall limit for each burst $n_\textrm{min}$ (from equation (\ref{eqn: true nmin})) is shown in Fig. \ref{hi} (top). The allowed values of circumburst density are to the right of the  cumulative distribution functions (CDFs). If we take the most constraining limit for each burst we obtain $f(<10^{-4})\lesssim0.16$. 

 Of course, none of these 19 SGRBs are those that have been labeled observationally hostless (observationally hostless SGRBs do not have measured redshifts as SGRB redshifts usually come from host associations). In fact, the singular burst that is consistent with having a density $n_{\textrm{min}}<10^{-6}\, \textrm{cm}^{-3}$ when using the upper limit on deceleration time is GRB 070724A which has a strong host association in a moderately star forming galaxy with a probability of chance coincidence $P_{cc}=0.002$ \citep{Kocevski2010}. This is of course completely consistent with our analysis, which only provides a lower limit on the true density. We note that there is no trend observed in these lower limits as a function of the host galaxy offset.

\par

\subsubsection{Sample without Redshifts}

\begin{figure*} 
\centering
\includegraphics[width=1.25 \columnwidth]{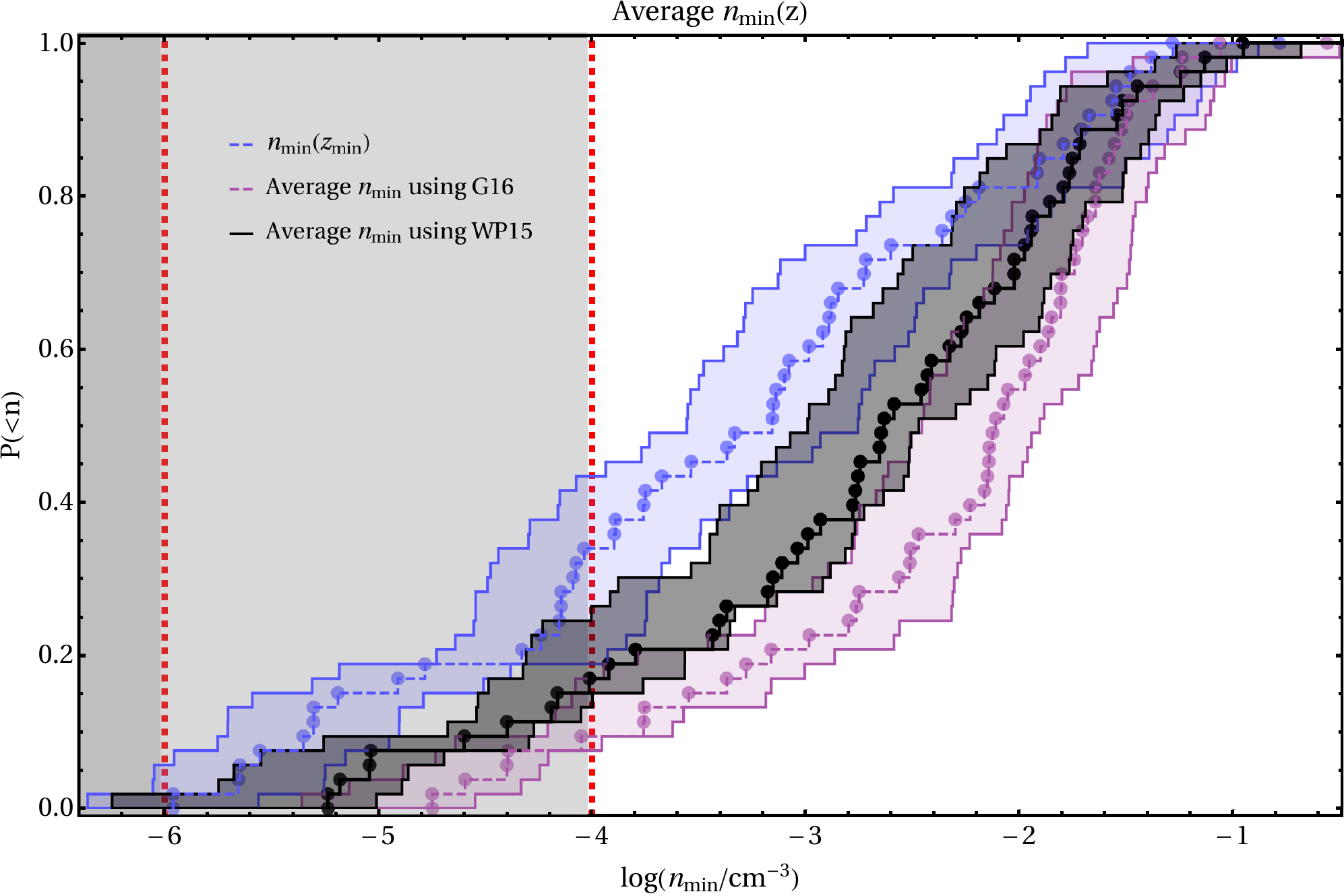}
\caption{
Cumulative distributions for $n_{\textrm{min}}$ calculated from equations (\ref{eqn: true nmin}) and (\ref{eqn: average nmin})  using $\Gamma=300$ and $\varepsilon_B=10^{-2}$. For equation (\ref{eqn: true nmin}), the redshift is chosen to minimize the lower limit $n_{\textrm{min}}(z_\textrm{min})$. We also show the average value of the lower limits $\langle n_\textrm{min}\rangle$, equation (\ref{eqn: average nmin}), using the redshift distributions of \citet{Wanderman2015} (WP15) and \citet{Ghirlanda2016} (G16). The dotted vertical lines, and shaded regions, represent densities consistent with $n_{\textrm{min}}\lesssim10^{-6}\, \textrm{cm}^{-3}$ and $n_{\textrm{min}}\lesssim10^{-4}\, \textrm{cm}^{-3}$, respectively.}
\label{hi2}
\end{figure*}

Here we demonstrate the results of our calculations for all 52 SGRBs in Table \ref{Table_SGRB_DATA}. The lower limits on circumburst density are presented in Fig. \ref{hi2}. First, we conservatively apply the redshift that minimizes the overall lower limit $n_\textrm{min}(z_\textrm{min})$ from equation (\ref{eqn: true nmin}). In addition, we explore the effect of redshift on our lower limits by averaging over the redshift distributions of WP15 and G16 using equation (\ref{eqn: average nmin}), leading to more realistic limits. 

We have shown in \S \ref{sec: density results 20 sample} that $\lesssim 16\%$ of SGRBs in our sample with redshift measurements are consistent with $n_\textrm{min}<10^{-4}$ cm$^{-3}$. We find a consistent result for our full sample of SGRBs, $f(<10^{-4})\lesssim 0.17$, from the average lower limit $\langle n_\textrm{min}\rangle$ using the redshift distribution $P(z)$ from WP15. We also identify for these limits that $f(<10^{-2})\lesssim 0.7$. In contrast, using $P(z)$ from G16 predicts less SGRBs at lower densities with $f(<10^{-2})\lesssim 0.5$. This has significant implications for the comparison with observations. If the G16 redshift distribution is correct then SGRBs must form very close to the center of their galaxies. Either way, we have shown that, regardless of the assumed redshift distribution, the central $1\sigma$ of the circumburst density distribution for SGRBs cannot vary by more than $\sim 2-3$ orders of magnitude as our lower limits provide a maximum on the scatter in density. We further explore the width of the SGRB density distribution in \S \ref{sec: 11-hour flux corr}.

\begin{figure} 
\centering
\includegraphics[width=\columnwidth]{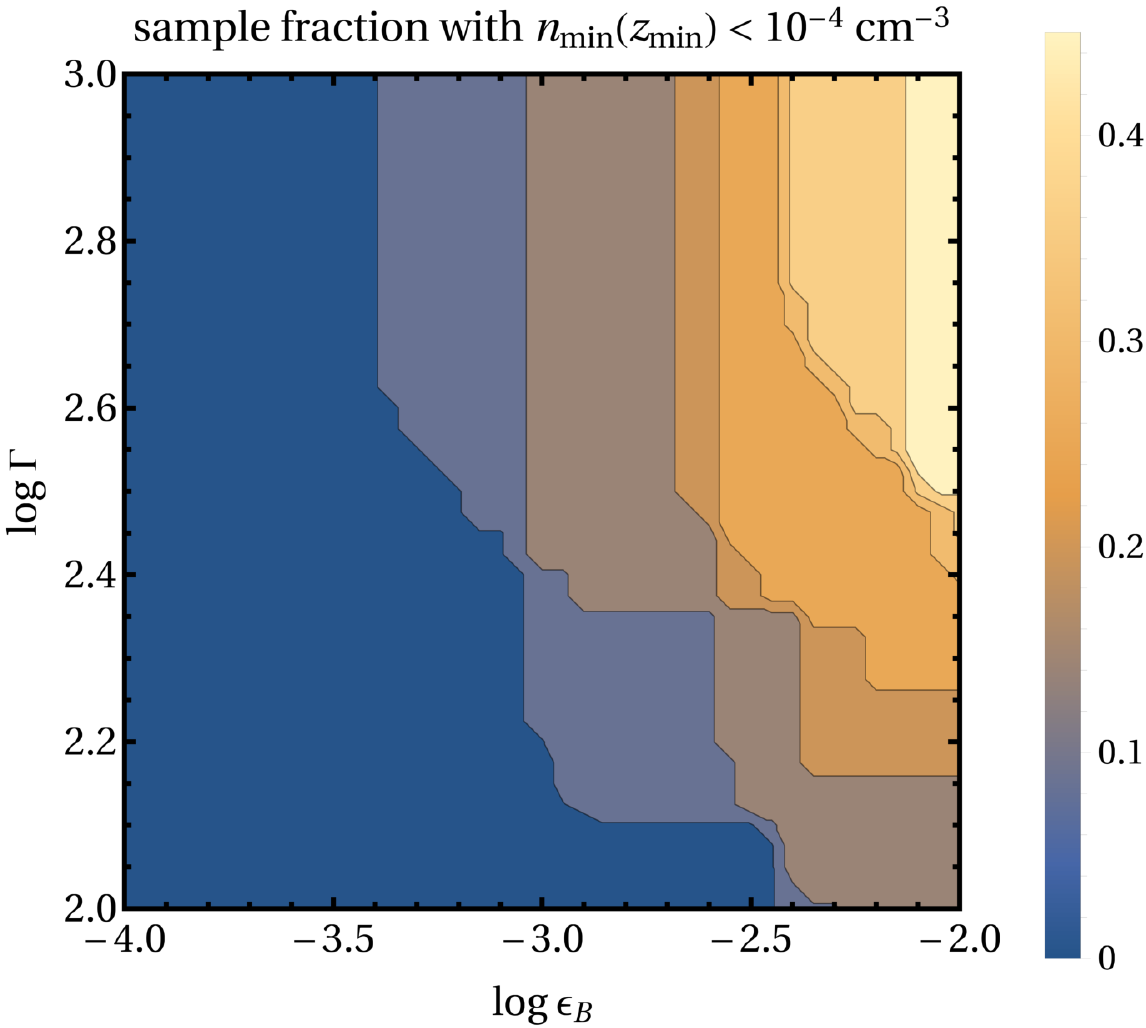}
\includegraphics[width=\columnwidth]{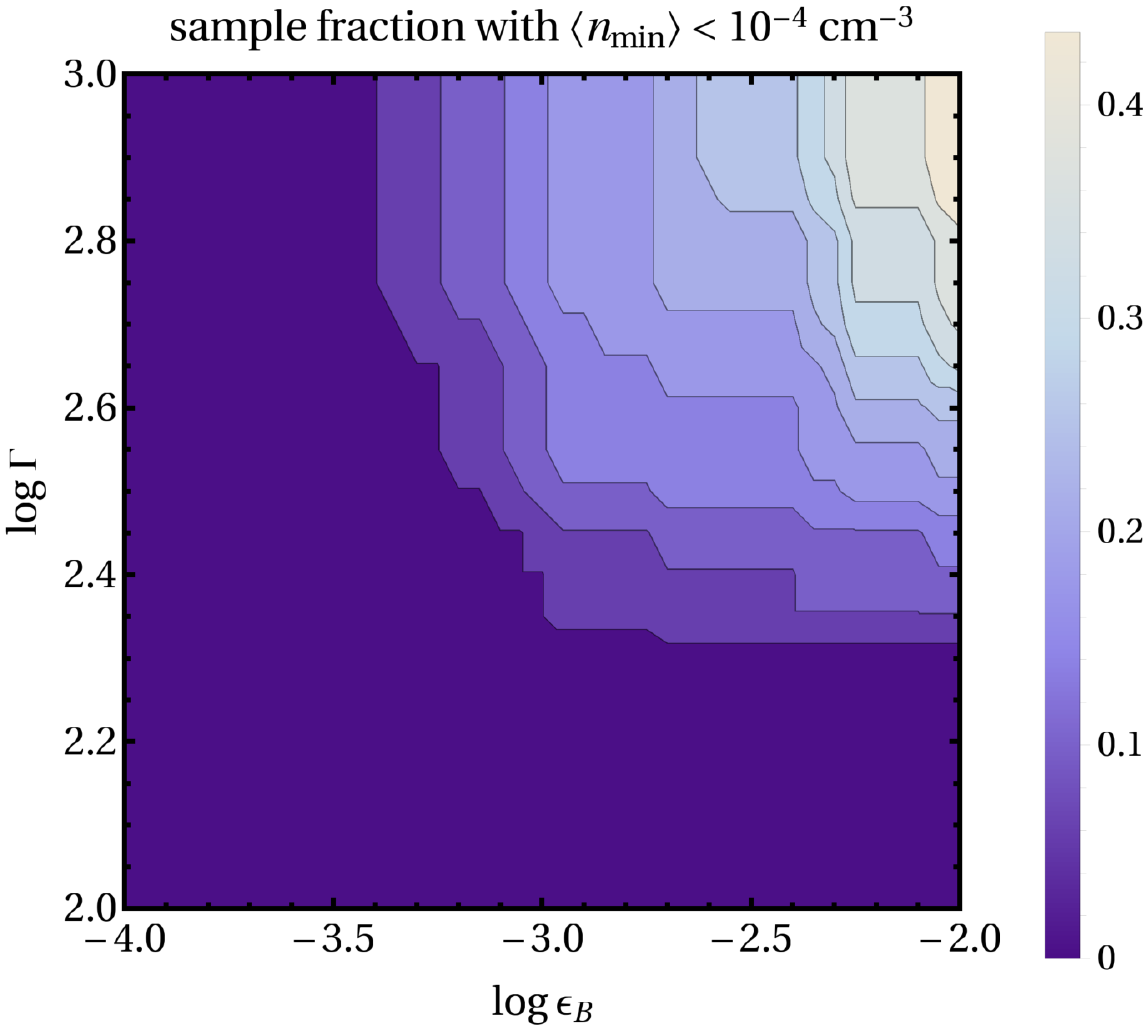}
\caption{Fraction of our overall sample of 52 SGRBs with $n_\textrm{min}(z_\textrm{min})<10^{-4}$ cm$^{-3}$ (Top) and average limits $\langle n_\textrm{min}\rangle<10^{-4}$ cm$^{-3}$ (Bottom) as a function of $\varepsilon_B$ and $\Gamma$. To calculate the average limits we used $P(z)$ from WP15, see equation (\ref{eqn: average nmin}). 
}
\label{fig: density plot phyhost}
\end{figure}

In Fig. \ref{fig: density plot phyhost}, we demonstrate the effect of varying $\varepsilon_B$ and $\Gamma$ on the fraction of bursts with $n_\textrm{min}(z_\textrm{min})<10^{-4}$ cm$^{-3}$ and average limits $\langle n_\textrm{min}\rangle<10^{-4}$ cm$^{-3}$, using $P(z)$ from WP15. This figure demonstrates that our choice of $\varepsilon_B=10^{-2}$ throughout is conservative as it leads to the highest fraction of bursts with density limits less than $10^{-4}$ cm$^{-3}$. Similarly, our choice of $\Gamma=300$ is conservative as for $\Gamma=100$ and $\varepsilon_B=10^{-2}$ we find $f(<10^{-4})\lesssim 0.15$ for $n_\textrm{min}(z_\textrm{min})$ (which is $\sim20\%$ less than for $\Gamma=300$). Averaging these lower limits over redshift (using the WP15 redshift distribution) provides a much more constraining limit, for $\Gamma=100$ and $\varepsilon_B=10^{-2}$ there are no bursts consistent with $\langle n_\textrm{min}\rangle\lesssim10^{-4}$ cm$^{-3}$.

\subsubsection{Application to LGRBs}

\begin{figure} 
\centering
\includegraphics[width=\columnwidth]{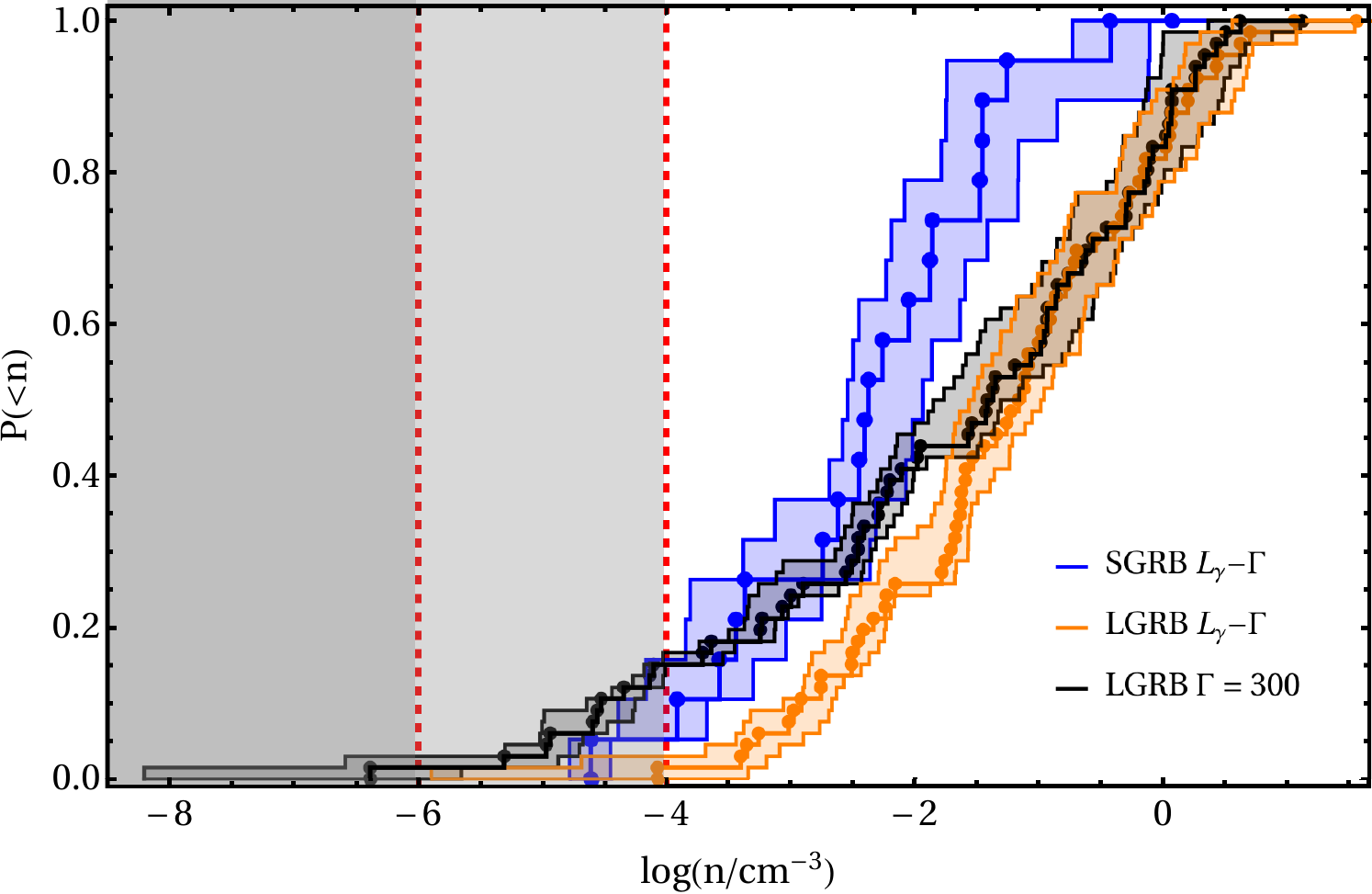}
\caption{Cumulative distribution functions for circumburst density estimates of 66 LGRBs based on optical afterglow peaks using  bulk Lorentz factor $\Gamma=300$ and the \citet{Lu2012} correlation for $L_\gamma-\Gamma$. Also shown is the CDF of lower limits on density $n_{t_o}$ for the 19 SGRBs with redshift using the $L_\gamma-\Gamma$ correlation (for $\varepsilon_B=10^{-2}$). The dotted vertical lines, and shaded regions, represent densities consistent with $n\lesssim10^{-6}\, \textrm{cm}^{-3}$ and $n\lesssim10^{-4}\, \textrm{cm}^{-3}$, respectively.}
\label{fig: LGRB CDF}
\end{figure}

As a comparison test case, we applied equation (\ref{decellim}) for a sample of 66 LGRBs from  \citet{Ghirlanda2017}, taking into account the requirement that $n_{t_o}<n_\textrm{crit}$. This ensures that electron's radiating in the optical are in the slow cooling regime. For these LGRBs, $n_\textrm{crit}$ is determined by solving for the density at which $\nu_c\approx 7\times 10^{14}$ Hz. This definition of $n_\textrm{crit}$ is used for LGRBs because the observations applied here were at optical frequencies \citep{Ghirlanda2017}. The cumulative distribution function presented in Fig. \ref{fig: LGRB CDF} represents {\it estimates} on the circumburst density for LGRBs (rather than just upper limits), assuming the $L_\gamma-\Gamma$ correlation from \citet{Lu2012} holds. This is because for our LGRB sample we have estimates of the peak times and fluxes, rather than just limits on those values as per the SGRB case. As these estimates are quite a bit lower than the expected densities for LGRBs, it is likely the correlation ($L_\gamma-\Gamma$) does not accurately reflect the bulk Lorentz factor for the entire sample of LGRBs, or else our assumption of the higher value of $\epsilon_B$ (i.e., $10^{-2}$) for all bursts was too conservative. We also perform the same calculation adopting a constant bulk Lorentz factor $\Gamma=300$ \citep{Ghirlanda2017}.

\citet{Soderberg2006} found that LGRB circumburst densities span a large range from $10^{-2}$ to $10$ cm$^{-3}$. We find a fraction ranging from $\sim0.1-3\%$ ($\sim1.5-15\%$) consistent with $n_{\textrm{min}}<10^{-6}
\,(10^{-4})\, \textrm{cm}^{-3}$ depending on bulk Lorentz factor. It is not expected for LGRBs to occur at such low densities as they are typically found near the half-light radius of their host galaxies \citep{Fruchter2006} and are expected to be associated with highly star forming regions \citep{Bloom2002,Hjorth2012}. In cases where LGRB hosts have not been detected to deep limits the host has been labeled as too faint for detection, as opposed to the GRB being hostless \citep{Hjorth2012,Tanvir2012}. If 3\% of LGRBs can be consistent with having $n_{\textrm{min}}<10^{-6}\, cm^{-3}$ when using $\Gamma=300$ it is plausible the same percent found for SGRBs is not due to their physically hostless nature but simply due to the conservative approach that we have adopted in this work. An alternative possibility is that the inferred LGRB densities are somewhat skewed due to our modelling based on a uniform density medium (as opposed to a wind external medium, $k=2$, which is appropriate for LGRBs).

\subsubsection{Lorentz Factor Correlations Applied to SGRBs}

We likewise calculate the circumburst density limits for SGRBs using the $\Gamma-L_\gamma$ correlation \citep{Lu2012}, see Fig. \ref{fig: LGRB CDF}. We choose this correlation (as opposed to $\Gamma-E_\gamma$) because \citet{Ghirlanda2009} found that SGRBs are consistent with the $E_p-L_\gamma$ correlations of LGRBs. We find an average lower limit that is an order of magnitude lower than the average for LGRBs. Of course, this is fully consistent with LGRBs and SGRBs occurring in similar density environments \citep{Nysewander2008}.

\subsection{SGRB Afterglow Peaks}
\label{sec: afterglow peaks obs}

In the standard afterglow model, the forward shock deceleration peaks of GRBs are preceded by $F_X\propto t^{3}$, and are followed by typical slopes of $F_X\propto t^{-1}$. In our sample, we identify two potential X-ray afterglow peaks in GRBs 060313 and 060801. We also consider GRB 090510, which was observed to have a deceleration peak in its \textit{Fermi}/LAT lightcurve \citep{Ackermann2010,Ghirlanda2010}.
The importance of the afterglow peaks are that the peak X-ray flux $F_{X,p}$ and peak time $t_p$ are no longer limits, but estimates. This allows us to estimate the circumburst density for these SGRBs if they are confirmed to have measurements of $F_{X,p}$ and $t_p$.
\par
In order to explore this possibility, we consider an X-ray flux with time dependence $F_X\propto t^{\alpha_X}$ prior to the afterglow peak. The X-ray slope of GRB 060313's lightcurve prior to the peak is  $\alpha_X=5.4^{+12.3}_{-4.3}$ given the $1\sigma$ error bars from the \textit{Swift}/XRT lightcurve Repository. For GRB 060801, we obtain $\alpha_X=2.0^{+6.0}_{-1.6}$. These slopes are consistent with $\alpha_X\sim3$, but the errors are quite large. Thus it is difficult to argue with a high level of certainty that these bursts have measured afterglow peaks. Here, we present the density estimates that would be obtained given that these were true measurements of afterglow peaks. 

\textit{GRB 060313} -- For GRB 060313, there is no measured redshift and we are, therefore, only able to apply equation (\ref{peaklumlimit}). This yields a density of $n = (6.4^{+10.5}_{-4.1})\times 10^{-5}\, \textrm{cm}^{-3}$ for $\Gamma=300$, $\varepsilon_B=10^{-2}$, and $\varepsilon_e=0.1$. Using a value of $\varepsilon_B=10^{-4}$ increases the estimate by two orders of magnitude. We note that GRB 060313 has a strong host association ($P_{cc}=3\times 10^{-3}$) in a faint galaxy ($m=26.4$ mag) \citep{FongBergerFox}. The offset between the SGRB location and the center of the host galaxy is $0.32\arcsec$ \citep{Berger2007}. 
\par
\textit{GRB 060801} -- GRB 060801 also has a host association with a late-type galaxy with $P_{cc}=0.02$ \citep[][and references there-in]{Fong2013}. The host's redshift of $z=1.13$ allows us to set equations (\ref{decellim}) and (\ref{peaklumlimit}) equal in order to obtain $\Gamma=224^{+71}_{-45}$ which yields a density estimate of  $n = (2.1^{+2.8}_{-1.8})\times 10^{-3}\, \textrm{cm}^{-3}$ using $\varepsilon_B=10^{-2}$ and $\varepsilon_e=0.1$. If the peak of the afterglow was in fact detected in X-rays then this is an estimate of the density and initial bulk Lorentz factor, but if not, then it is a lower limit. 
\par
\textit{GRB 090510} -- GRB 090510 was the first SGRB detected in the GeV range by \textit{Fermi}/LAT. A deceleration peak of $t_p=0.2$ s was measured from the LAT lightcurve \citep{Ackermann2010, Ghirlanda2010}. Due to the fact that the peak of the afterglow was likely detected in the fast cooling regime (for typical burst parameters) the environment is less constraining and we cannot use the peak luminosity to set relevant limits on the density. The method of equation (\ref{peaklumlimit}) is likely not applicable to afterglow peaks detected by \textit{Fermi}, but applies to those detected in X-rays (e.g., \textit{Swift}/XRT detections). If the emission is assumed to be synchrotron radiated at the forward shock then the deceleration peak $t_p=0.2$ s leads to $\Gamma\approx 2000$ for a circumburst density $n\approx 1$ cm$^{-3}$ \citep{Ghirlanda2010, Ackermann2010}. We note that GRB 090510 has a host galaxy at an offset of $1.2\arcsec$ \citep[9.4 kpc at $z=0.903$,][]{Rau2009,McBreen2010}.  

\subsection{Application to Observationally Hostless SGRBs}
\label{sec_hostlesslimits}

\begin{table*}
\centering
\caption{The $3sigma$ point source limits on coincident host galaxies for the eight observationally hostless SGRBs. References: [1] \citet{FongBergerFox} [2] \citet{Berger2010a}  [3] \citet{FongBerger2013} [4] \citet{Perley2008} [5] \citet{Tunnicliffe2014} [6] \citet{Fong2013} [7] \citet{Fong2012} [8] \citet{Perley2009} [9] \citet{Rowlinson2010} } 
\begin{tabular}{|c|c|c|c|c|}
\hline
\hline
GRB &  Apparent Magnitude (AB) &  & & Ref. \\
\hline
\hline
   \multicolumn{5}{c}{hostless SGRBs in our sample}  \\
\hline
\hline
061201 &  >26.2 (HST: F160W)& >26 (HST: F814W)& >25.5 (HST: F606W) & [1]-[3] \\

070809 & >26.2 (HST: F160W) &  >26.3 (Keck I (+LRIS):g)& >25.4 (Magellan LDSS3: r) & [2]-[4] \\

091109B& >25 (HST: F160W)& >25.8 (VLT FORS2: R)&>22.23 (VLT HAWK-I: K) & [3],[5],[6] \\

110112A &  >26.2 (Gemini GMOS-N: i) & >25.5 (Magellan LDSS3: r)& >24.7 (Magellan LDSS3: i) & [6] \\

111020A & >25.4 (Gemini GMOS-S: i) & >24.03 (VLT FORS2: R)& -- & [5],[7] \\
\hline
\hline
\multicolumn{5}{c}{other hostless SGRBs}  \\
\hline
\hline
080503 &   >28.5 (HST: F606W) & >25.7 (HST: F606W)& -- & [2],[8] \\
090305A &   >25.9 (VLT FORS2: R) & >25.69 (Gemini GMOS-S: r)& >25.6 (Magellan LDSS3: r) & [2],[5] \\
090515 &  >26.5 (Gemini GMOS-N: r) & -- & -- & [2],[9] \\

\hline
\end{tabular}
\label{Table_Hostless_Bursts}
\end{table*}
\par
In Table \ref{Table_Hostless_Bursts}, we present the apparent magnitude limits on coincident host galaxies for the observationally hostless SGRBs. There are no coincident hosts to these deep limits which has led to their observationally hostless classification. Based on our calculated lower limits on circumburst density none of these are physically hostless (i.e., $n_\textrm{min}\gtrsim 10^{-4}$ cm$^{-3}$), which suggests they reside within host galaxies at moderate to high redshifts where they are not detected by these follow-up observations in the optical and NIR. 
\par
We remind the reader that observationally hostless bursts can arise from two main avenues: (i) there are multiple galaxies with similar probabilities of chance coincidence due to a crowded field or (ii) the galaxy with the lowest probability of chance coincidence has $P_{\textrm{cutoff}}<P_{cc}$, where $P_{\textrm{cutoff}}$ is the highest probability leading to a strong host association. The chance probability cutoff varies in the literature with typical values: 0.1 \citep{Bloom2002,Berger2010a,Blanchard2016}, 0.05 \citep{Fong2013}, or 0.01 \citet{Tunnicliffe2014}. We note that for a Milky Way-like spiral galaxy a $P_{cc}>0.1$ occurs at $\sim 20$ kpc from the galaxy's center at $z=1$ \citep{Church2011,Tunnicliffe2014}, which is still well within the host's virial radius\footnote{The Milky Way's virial radius is $\sim$ 200 kpc.}.

\begin{figure} 
\centering
\includegraphics[width=\columnwidth]{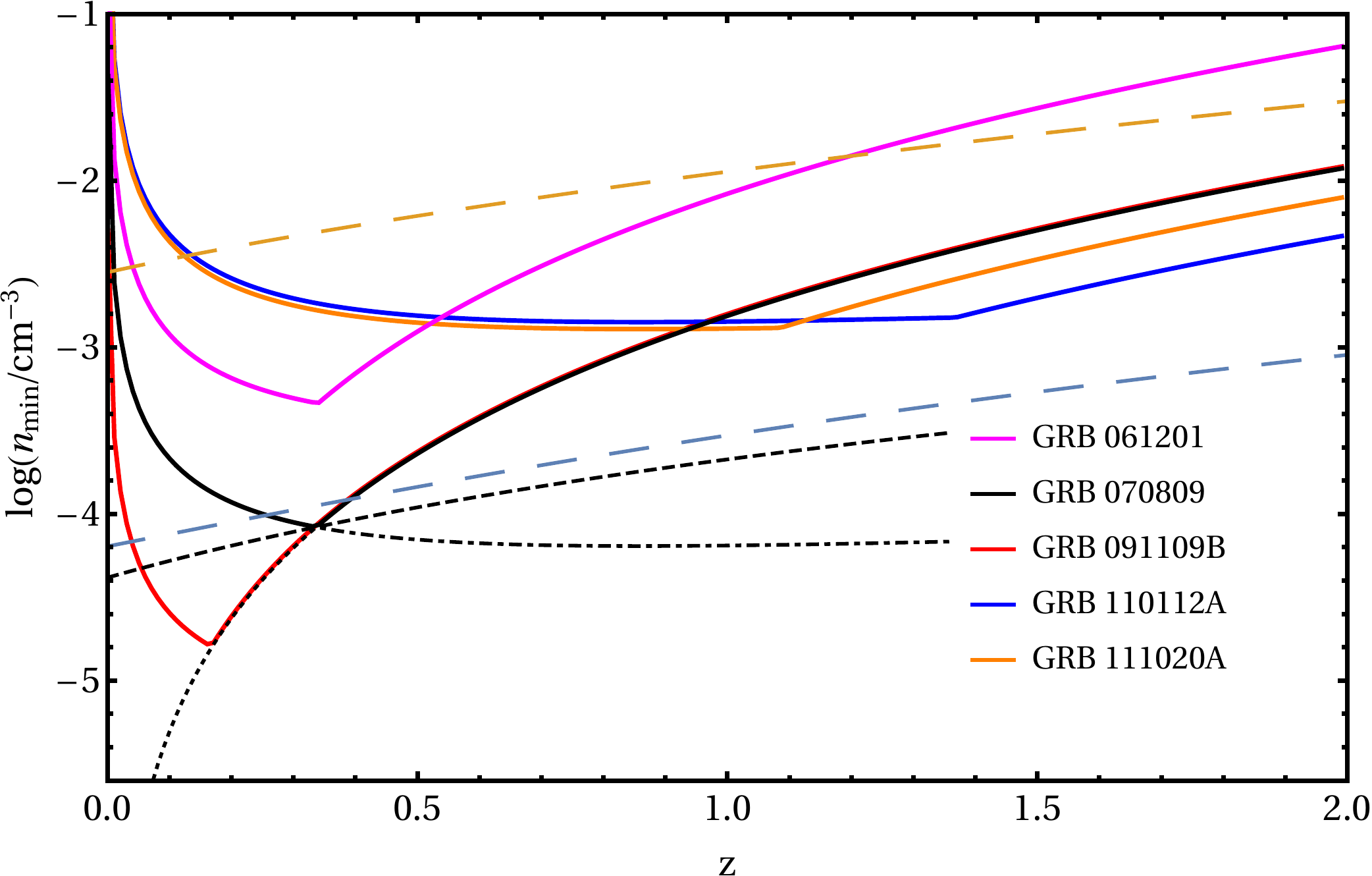}
\caption{Most constraining limit on circumburst density for the five observationally hostless bursts as a function of redshift. The error on the density lower limit for each burst is $\sigma_{\log n_{\textrm{min}}}\approx 0.4$. For GRB 070809, we show the lower limits calculated using the deceleration time (equation (\ref{decellim}), black medium dashed), peak flux (equation (\ref{peaklumlimit}), black short dashed), and combined limit (equation (\ref{combinedlimit}), black dot dashed) as an example. Also shown is the density at the virial radius (blue long dashed lines) and at $5r_e$ (yellow long dashed lines) for a typical SGRB host galaxy, $M_\textrm{vir}=10^{12} M_\odot$ \citep{Behroozi2014}.}
\label{obshostlesslimitsboth}
\end{figure}

In order to distinguish the observationally hostless bursts from being consistent with having occurred in $n_{\textrm{min}}<n_\textrm{vir}$, we calculate limits on the density using equation (\ref{combinedlimit}) as a function of redshift. The results of this calculation are shown in Fig. \ref{obshostlesslimitsboth}. We identify that none of the five observationally hostless bursts in our sample are physically hostless (at the redshift for the lowest chance probability galaxy). This is individually discussed in Appendix \ref{Appendix: obs hostless} for these bursts (GRBs 061201, 070809, 091109B, 110112A, and 111020A).

Three of the observationally hostless SGRBs: 080503A, 090305A, and 090515 have only optical detections that can be securely associated with the afterglow phase (i.e., excluding possible contamination of the X-ray lightcurve from HLE, see \S \ref{sec: obs-hostless observations section}). Our methodology discussed in \S \ref{sec: methods} can be easily adopted to the optical band. However, in practice, the optical detections in these bursts are at very late times (respectively, $t_o\approx 2000$ s, $4000$ s, and $6000$ s as opposed to $\langle t_o\rangle = 200$ s which is the characteristic value for SGRBs in our X-ray sample). Since our method relies on detections as close as possible to the time of the deceleration peak, these optical measurements, therefore, do not provide any meaningful constraints on the external density. However, we do mention the host associations of these three bursts in Appendix \ref{Appendix: obs hostless}.

\subsection{11-Hour Flux Correlation}
\label{sec: 11-hour flux corr}

\begin{figure} 
\centering
\includegraphics[width=\columnwidth]{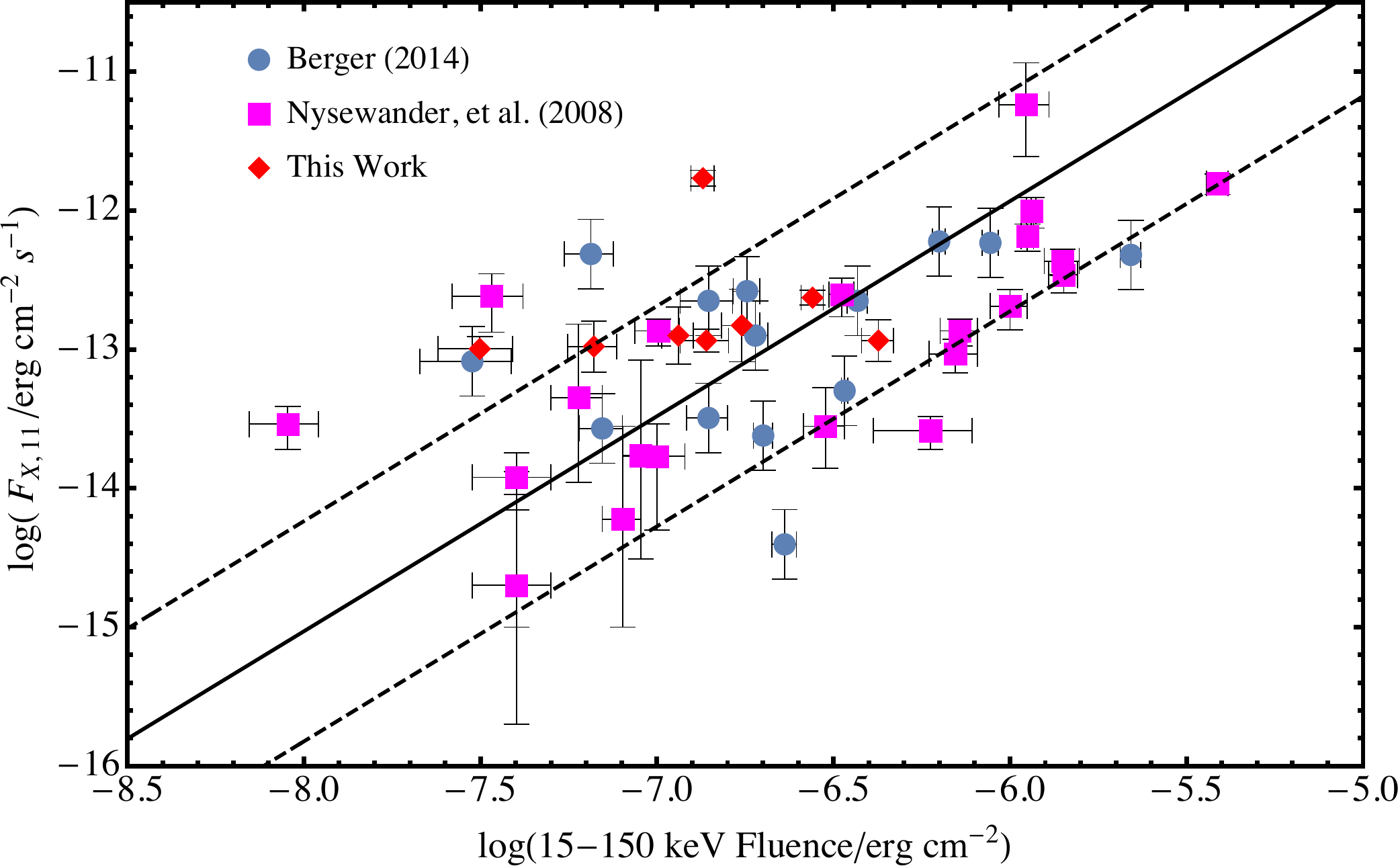}
\caption{The X-ray flux ($0.3-10$ keV) at 11-hours versus the BAT fluence (15-150 keV) for SGRBs from \citet{Nysewander2008} (squares), \citet{Berger2013} (circles), and for bursts added in this work (diamonds). The error on bursts from \citet{Berger2013} is taken to be the average error from \citet{Nysewander2008}. Also shown is the best fit linear correlation and $1\sigma$ standard error.}
\label{fig: 11-hr corr}
\end{figure}

We investigate the correlation between 11-hour X-ray flux and BAT fluence presented by \citet{Nysewander2008} for a sample of 37 SGRBs. 
We use a sample of 44 SGRBs with X-ray detections at 11-hours, shown in Fig. \ref{fig: 11-hr corr}. To test the correlation, we apply a bootstrap algorithm with pairwise replacement for $X=\log \phi_\gamma$ and $Y=\log F_{X,11}$ to obtain a Pearson correlation coefficient $r=0.56^{+0.15}_{-0.20}$ (95\% CI). Using a permutation test, we rule out the null hypothesis of no correlation $\rho=0$ with $p-$value $p=3.6\times 10^{-4}$, where $\rho$ is the true correlation coefficient and $r$ is the sample correlation coefficient. These tests indicate that a correlation between $F_{X,11}$ and $\phi_\gamma$ exists (Fig. \ref{fig: 11-hr corr}). We identity a best fit correlation of 
\begin{align}
    \log F_{X,11}= (1.55 \pm 0.33) \log \phi_\gamma - (2.64\pm2.19)
\end{align}
The expected slope of this correlation in the slow cooling regime is $ \log F_{X,11}\propto 1.3 \log \phi_\gamma$, whereas in the fast cooling regime it is $ \log F_{X,11}\propto 1.05 \log \phi_\gamma$. We find that the correlation is in agreement with the expected slope for the slow cooling regime ($\nu_m<\nu<\nu_c$) at the $1\sigma$ level and with the fast cooling slope ($\nu>\nu_c,\nu_m$) at $2\sigma$.

Some GRB afterglow modelling attempts have suggested that the external densities of SGRBs may be distributed over 6 orders of magnitude with typical values ranging from $10^{-6}-1\mbox{ cm}^{-3}$ \citep{Fong2015}. In this case the relevant relation between $F_X$ and $\phi_{\gamma}$ for most bursts would be the slow cooling relation, equation (\ref{eqn: 11-hour flux SC}), and the scatter in density would be $\sigma_{\log( F_{x,11}/\phi_\gamma)}\approx 3$ (taking, as an illustration, a uniform distribution in log-space). However, in practice $\sigma_{\log( F_{x,11}/\phi_\gamma)}=0.64$ which is much smaller. The width of this correlation can be used to estimate the median value and the scatter in density. We use Monte Carlo simulations to determine the circumburst density distribution required to recreate the observed scatter in $\log F_{X,11}/\phi_\gamma$, see Appendix \ref{Appendix: Monte Carlo} for details. 
 
 We first perform these simulations with a fixed density to test the scatter in the observed correlation due to changing $\langle \log \varepsilon_B\rangle$ and $\langle \log n\rangle$. We find that for fixed density and  $\sigma_{\log \varepsilon_B}=1$ it is possible to reproduce the observed correlation for a number of combinations, e.g., $\langle \log n \rangle=\{-1,-2,-3\}$ and $\langle \log \varepsilon_B\rangle=\{-2,-3,-4\}$, among others. Therefore, given the scatter in $\varepsilon_B$ found from afterglow modeling \citep{Santana2014,Zhang2015}, the density distribution for SGRBs is not required to have a large width. We note that for fixed density $\log n=-2$ and fixed $\varepsilon_B=10^{-4}$ the scatter in other parameters ($z$, $E_{p,\textrm{source}}$, $E_\gamma$, and $\varepsilon_e$) produce at most half of the observed width of the $F_{x,11}/\phi_\gamma$ distribution (for other fixed values of density and $\varepsilon_B$ there is less scatter as a higher percentage of bursts have electrons radiating at $\nu_X\sim 1$ keV in the fast cooling regime). In order to produce the observed scatter we require either $\sigma_{\log \varepsilon_B}\gtrsim 1 $ for fixed density, $\sigma_{\log n}\gtrsim 2$ for fixed $\varepsilon_B$, or, e.g., $\sigma_{\log \varepsilon_B}\gtrsim 0.75$ and $\sigma_{\log n}\gtrsim 0.5$. These results are sensitive to changes in $\langle \log \varepsilon_B\rangle$ and $\langle \log n\rangle$.

We perform the same analysis for different density distributions. The results are tabulated in Table \ref{tab: scatter sim results}, and represent only a sample of the possible distributions of $n$ and $\varepsilon_B$ capable of reproducing observations. The correlation is not strong enough that we can state conclusively from the correlation alone that the density distribution must be narrow as the intrinsic distribution of SGRB parameters (e.g., $\varepsilon_B$) is not well known. Under the assumption that there is only scatter in the density, we cannot rule out a wide distribution of densities ($\sigma_{\log n}\gtrsim 0.75$) though we find a population of low density bursts is not required. As mentioned above, the latter possibility is somewhat disfavoured given the comparable correlation width observed in LGRBs, which is expected to be dominated by $\epsilon_B$ rather than external density variance.

\begin{table}
 	\centering
 	\caption{Input parameters and results of the Monte Carlo simulations used to determine the distributions of $n$ and $\varepsilon_B$ that reproduce the observed scatter in $\log(F_{X,11}/\phi_\gamma)$. These simulations are performed using two different luminosity functions (LF) from \citet{Wanderman2015} (WP15) and \citet{Ghirlanda2016} (G16). We track the fraction of detected bursts which have X-ray radiation emitted by electrons in the slow cooling (SC) regime. }
 	\label{tab: scatter sim results} 
 	\begin{tabular}{lcccc}
    \hline
    \hline
  LF & $\langle \log n\rangle\pm \sigma_{\log n}$ & $\langle \log \varepsilon_B\rangle\pm \sigma_{\log \varepsilon_B}$ & $\sigma_{\log\Big(\frac{F_{X,11}}{\phi_\gamma}\Big)}$ & \% SC \\
    \hline
    \hline
   G16   & $-2\pm1$ & $-3\pm1 $ & 0.63 & 57 \\
   G16   & $-2\pm1$ & $-4\pm1 $ & 0.66 & 74 \\
   G16   & $-3\pm0.5$ & $-2\pm1 $ & 0.65 &  65 \\
   G16   & $-3\pm0.5$ & $-3\pm0.75 $ & 0.63& 94  \\
G16   & $-4\pm0.5$ & $-2\pm0.75 $ & 0.67& 94  \\
    \hline
  WP15   & $-1\pm1$ & $-3\pm1 $ & 0.63 & 55 \\
  WP15   & $-1\pm1$ & $-4\pm1 $ & 0.67 & 71 \\
  WP15   & $-2\pm0.5$ & $-3\pm0.75 $ & 0.62 & 90   \\
  WP15   & $-2\pm0.75$ & $-4\pm0.75 $ & 0.64 & 98   \\
  WP15   & $-3\pm0.5$ & $-2\pm 0.75 $ & 0.66 & 93 \\
    \hline
    \end{tabular}
\end{table}

\subsection{Constraints on Offset Distribution}

\begin{figure} 
\centering
\includegraphics[width=\columnwidth]{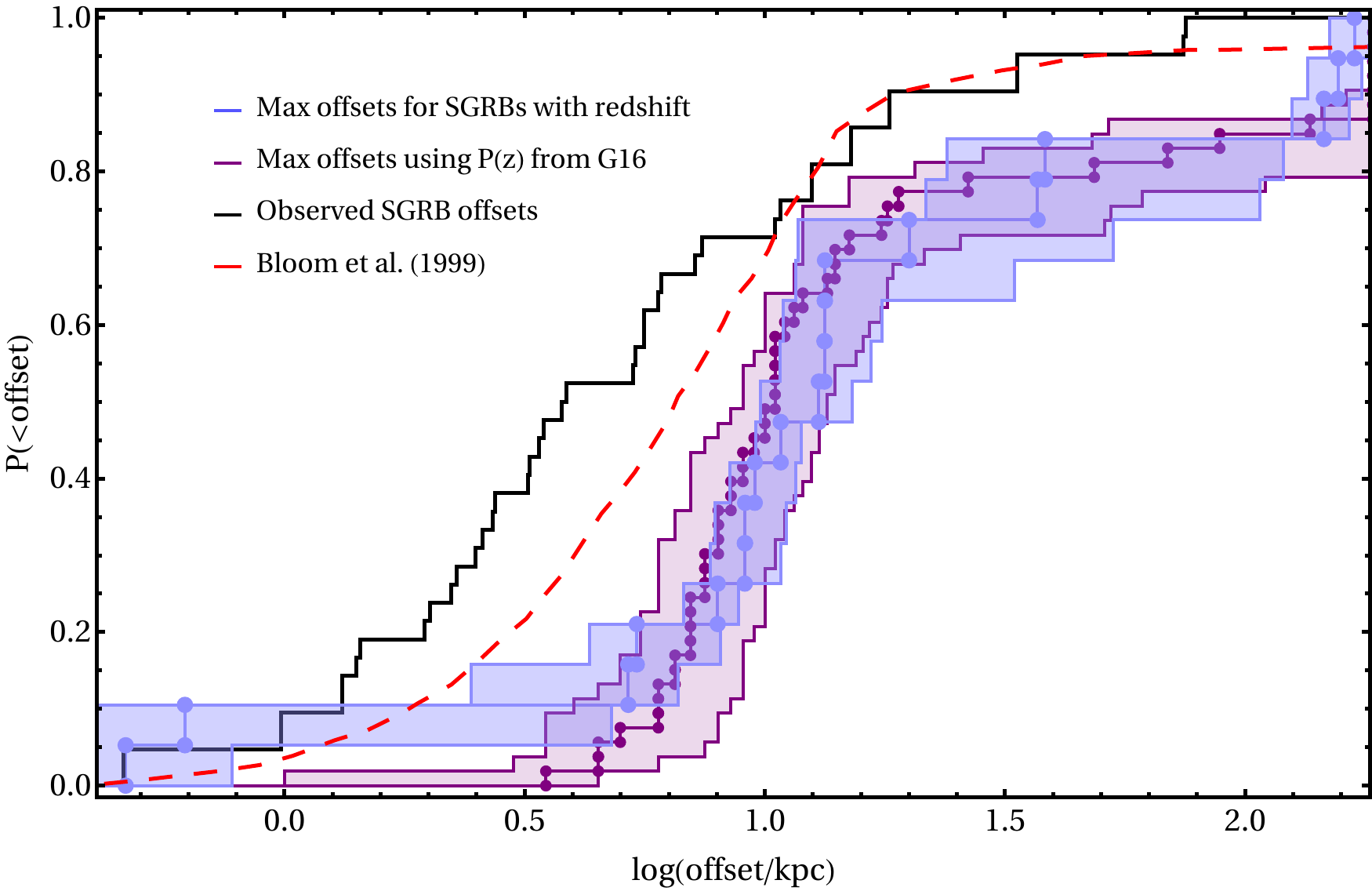}
\caption{Cumulative distribution demonstrating the fraction of systems occurring at a given physical offset (in kpc). We present upper limits on physical offset for our sample of 19 SGRBs with redshift and for the average lower limits $\langle n_\textrm{min}\rangle$ for our full sample (52 SGRBs) computed using $P(z)$ from G16. These limits are computed assuming the gas profile described in \S \ref{sec: gas density profile}. 
We compare these upper limits to observed SGRB offsets \citep{FongBerger2013} and to theoretical expectations \citep{Bloom1999}.
}
\label{fig: max offsets}
\end{figure}

In Fig.  \ref{fig: max offsets}, we present upper limits on the offset distribution for SGRBs based on the lower limits on circumburst density shown in Fig. \ref{hi} for 19 SGRBs with redshift measurements. We compute these upper limits using the gas density profile described in \S \ref{sec: gas density profile}. In this calculation, we apply estimates of host galaxy stellar mass for 14 (70\%) SGRBs found in the literature \citep{Leibler2010,FongBerger2013}. For those without estimates of stellar mass (30\%) we adopt the mean of the observed distribution $\log(M_*/M_\odot)\approx 9.9$. We cannot accurately constrain the offset for 3 (15\%) of these SGRBs because the lower limits on density are smaller than the density at the virial radius. As such, for the purpose of Fig. \ref{fig: max offsets}, we apply the virial radius as the maximal offset ($r_\textrm{max}\sim r_\textrm{vir}$). 

We find that the distribution of upper limits on offset (merger radius) are consistent with both the observed distribution \citep{Berger2010a,FongBerger2013} and theoretical predictions based on population synthesis modeling \citep{Bloom1999}. This demonstrates that the majority of SGRBs ($\gtrsim 84\%$) are constrained to within their host's virial radius and are not physically hostless. For the mean observed SGRB host mass the density at $r_e$ and $5r_e$ of $10^{-1}$ cm$^{-3}$ and $10^{-2}$ cm$^{-3}$ predicts that a minimum of 10\% and 45\% of the progenitor systems merged within $r_e$ and $5r_e$, respectively, see Fig. \ref{hi} (Top). Therefore, in this sample of SGRBs a significant portion could have occurred within these radii and close to their galactic centers. 

As a way to extend this calculation to our full sample, we calculate a physical offset averaged over the redshift distributions $P(z)$ from WP15 and G16 using the density limits $\langle n_\textrm{min}\rangle$ from Fig. \ref{hi2}. For both redshift distributions, we calculate $\langle n(r)\rangle = \int P(z) \,n(r|z, M_*)\, dz$, where we have adopted the average observed SGRB host galaxy stellar mass for each GRB. The offsets are then inferred directly from the density limits shown in Fig. \ref{hi2} for all 52 bursts. We find that the offsets computed for WP15 and G16 are consistent, but that the WP15 offsets are more sensitive to changes in the lower limit on density at large offsets. We present the average offset distribution using the redshift distribution from G16 in Fig. \ref{fig: max offsets}. We note that these offsets are upper limits and not estimates of physical offsets. For bursts with $\langle n_\textrm{min}\rangle \lesssim \langle n(r_\textrm{vir})\rangle$ we follow the same procedure described above and adopt $r_\textrm{max}\sim r_\textrm{vir}$. We find that both distributions of offsets (WP15 and G16) are consistent with the observed offset distribution. However, the results are sensitive to changes in the gas density profile at large radii. Adopting a steeper gas profile would lead to the inferred offsets for density limits from G16 becoming inconsistent with the observed offset distribution. 

\section{Conclusions}
\label{sec: conclusions}

In this paper, we have applied X-ray afterglow observations as a way to constrain the circumburst density distribution of the \textit{Swift} population of SGRBs. We utilized upper limits on the deceleration time and lower limits on the peak X-ray flux (see \S \ref{sec: methods}) to set lower limits on circumburst density for a sample of 52 SGRBs (Table \ref{Table_SGRB_DATA}). We summarize the results in Table \ref{Table_Full_Results}. Using the most conservative limit, $n_\textrm{min}(z_\textrm{min})$, we find no SGRB environment consistent with $n_\textrm{min}<10^{-6}$ cm$^{-3}$ (see Fig. \ref{hi2}). We identify that the fraction with density $n_\textrm{min}<10^{-4}$ cm$^{-3}$ is $f(<10^{-4})\lesssim 0.16$ (see Fig. \ref{hi}). We have defined a physically hostless SGRB as having occurred outside of its host galaxy's virial radius. By assuming a typical density at the virial radius of $n_\textrm{vir}\sim10^{-4}$ cm$^{-3}$, we are able to constrain the physically hostless fraction to $\lesssim 16\%$. This result is highly conservative (we have assumed $\Gamma=300$ and $\varepsilon_B=10^{-2}$, see \S \ref{sec: lower limits methods sec} for a discussion); taking, for example, more realistic values of $\Gamma=100$ and $\varepsilon_B=10^{-4}$, the fraction of physically hostless SGRBs is negligible (see Figure \ref{fig: density plot phyhost}). We note, however, that there is some tension between the results of this work and previous estimates on the density of SGRBs from broadband afterglow modeling \citep[e.g.,][]{Panaitescu2001,Fong2015}. This can be explained by the degeneracy in $\varepsilon_B$ and $n$ that exists in broadband afterglow modelling, specifically when the majority of observations are beneath the cooling break. In addition, these earlier works do not include SSC cooling effects on the synchrotron spectrum, e.g., equations (\ref{eqn: Yparam globally slow cooling}) and (\ref{eqn: Yparam globally fast cooling}). The latter is an unavoidable mechanism in the physical emission region (that adds no additional parameters to the modelling) that is typically ignored for the sake of simplicity. It has been shown that not including SSC effects can have a large effect on the inferred afterglow parameters \citep[e.g.,][]{Sari2001,Nakar2009}.

The analysis employed here not only constrains the fraction of bursts with low external densities, but also those with high densities. We find $\gtrsim30\%$ of our sample are constrained to have occurred in densities larger than $10^{-2}$ cm$^{-3}$ (assuming the redshift distribution from WP15). SGRBs at these densities are strongly over represented as electromagnetic counterparts to gravitational wave (GW) detections of BNS mergers as events in high densities have more luminous off-axis afterglow peaks \citep{Duque2019}. The detected fraction of high density events can provide strong constraints on the intrinsic population of SGRBs with high circumburst density, when taking into account these selection effects \citep{Duque2019}. By adopting a high density of $n_2=10^{-1}$ cm$^{-3}$ and low density $n_1=10^{-3}$ cm$^{-3}$ we set a lower limit on the observed fraction of electromagnetic counterparts with high densities $f_\textrm{HD}^\textrm{obs}$ given an intrinsic fraction of high density events $f_\textrm{HD}\gtrsim0.08$ found within this work. Assuming $N=40$ off-axis afterglow counterparts to GW detections, we find $f_\textrm{HD}^\textrm{obs}\gtrsim 0.175$ (0.05) for the energy distribution function from WP15 (G16) at the 95\% confidence level.

The large majority of SGRBs with redshift have their redshift determined through host galaxy association, as opposed to a spectroscopic measurement of the afterglow. The observed redshift distribution of SGRBs is thus highly dependent on the method used to assign a likelihood probability to potential host galaxies (this will be explored in future work). Due to the low redshift completeness and uncertainty in the merger delay time distribution, redshift distributions vary greatly in the literature and have significant implications for the comparison with observations. In Fig. \ref{hi2}, we employ the redshift distributions from WP15 and G16 to determine an average lower limit, $\langle n_\textrm{min}\rangle$. G16 predicts more SGRBs forming at higher redshifts with a peak in probability around $z\approx 2$, whereas WP15 predicts a peak of the SGRB redshift distribution around $z\approx 0.9$ with almost no events at $z>2$. These two distributions produce entirely different limits on density for our sample of bursts. The G16 redshift distribution requires that SGRBs must form very close to the center of their galaxies, although the maximum offsets are consistent with the observed offset distribution. This result depends on our choice of gas density profile; for profiles that are less flat at large radii the inferred offsets would become inconsistent with the observed distribution (i.e., more SGRBs are smaller offsets).

The fraction of physically hostless SGRBs is also highly dependent on the assumed gas profile (especially at large radii). Changing slightly our point of view, for SGRBs that are well constrained to be significantly kicked from their host galaxies (due, for example, to very deep optical follow-up), the method described in this work can use the afterglow of highly kicked SGRBs to constrain the galaxy density profile at large radii where other methods are typically ineffective \citep{Li2020}. We argue that either typical SGRBs occur very close to their galactic centers or that the gas density profile is quite flat out to large radii (e.g., the MB profile assumed in this work, see \S \ref{sec: gas density profile}) in order to have a large ($\sim 16\%$) physically hostless fraction.

In our sample of SGRBs, there are five bursts that have been labeled as observationally hostless in the literature, which means they have no coincident host galaxy and other nearby galaxies have large probabilities of chance coincidence \citep{Perley2009,FongBergerFox,Rowlinson2010,Berger2010a,Fong2012,Fong2013,FongBerger2013,Tunnicliffe2014}. We find that, although these SGRBs lack strong host galaxy associations, they are inconsistent with having occurred outside of their galaxy's virial radius. This is demonstrated clearly in Fig. \ref{obshostlesslimitsboth}, where we show that the most constraining limit as a function of redshift restricts these SGRBs to $n_\textrm{min}>10^{-4}$  cm$^{-3}$, which is the typical density at the virial radius for a SGRB host galaxy. Therefore, these observationally hostless bursts (GRBs 061201, 070809, 091109B, 110112A, and 111020A) are inconsistent with being physically hostless, see Appendix \ref{Appendix: obs hostless} for details. 

We have excluded 3/8 of the observationally hostless GRBs from our sample due to our selection criteria. However, we note that these events, due to their faint afterglows, are the most likely candidate physically hostless GRBs. In the case of GRB 080503, the early X-ray lightcurve shows HLE \citep{Perley2009}, which excludes it from our analysis. This GRB has the deepest limits on a coincident host galaxy \citep[HST/F606W$>28.5$ mag,][]{Perley2009}; although there is a potential galaxy $\sim 0.8\arcsec$ from the GRB position (see Appendix \ref{Appendix: obs hostless}). The dim afterglow for this event hints at a very low density environment $n\lesssim 10^{-5}$ cm$^{-3}$ (or a low value of $\varepsilon_B\lesssim 10^{-4}$), see \S 3.2 of \citet{Perley2009} for a detailed discussion. If the late-time rebrightening of the afterglow (in the optical and X-rays) at $\sim 1$ day is interpreted as the deceleration time \citep{Perley2009} then GRB 080503 must have had $\Gamma\lesssim 24$ in order to be consistent with $n>10^{-4}$ cm$^{-3}$. These Lorentz factors are at the low end of predicted values for SGRBs, with typical values of $\Gamma\gtrsim 10-50$ \citep{Nakar2007}.
This event suggests that it is possible that physically hostless SGRBs exist, but that their identification is difficult. If GRB 080503 is physically hostless, it is consistent (being 1/8 of the observationally hostless GRBs) with our upper limit of $<16\%$. This demonstrates that our methods can be utilized to determine the sub-sample of observationally hostless SGRBs that are potentially truly physically hostless.

As mentioned above, our lower limits on the densities suggest that, at least in the majority of observationally hostless SGRBs, we do not expect the BNS mergers to have occurred outside their galactic halos. One possible caveat, is if the birth galaxy of the BNS was part of a rich galaxy cluster \citep{Niino2008,Salvaterra2010}. In this scenario, a BNS may escape its host while still merging within the relatively dense (compared to the intergalactic medium) intra-cluster medium, the density of which can be as large as $10^{-3}\mbox{ cm}^{-3}$. Such a density is consistent with a significantly larger fraction of our SGRB population ($30-50\%$, see Figure \ref{hi2}). We note however that \citet{Berger2007a} have found that only $5-20\%$  of observed SGRBs are consistent with having taken place in a dense galaxy cluster. Furthermore, there is no evidence that any of the potential hosts of observationally hostless SGRBs reside in such clusters.
 
We therefore consider two alternative explanations for why observationally hostless SGRBs lack strong host associations. The first is that they belong to low redshift hosts where the NS natal kick has forced them to a location where they are within their birth galaxy's virial radius but outside of a region where there is a high probability of association for that host (i.e., the offset is large enough that the probability of random sky alignment with a galaxy is high). This is due to the fact that at low redshifts, moderate physical offsets appear as large angular sizes on the sky. This could lead to multiple potential hosts with similar probabilities of chance coincidence (see Appendix \ref{Appendix: obs hostless}), which would mask the true host association. The second possible explanation is that they belong to a moderate to high redshift population where the host galaxy's light is not uncovered by observational follow-up (e.g., galaxies fainter than $\sim 26$ mag). \citet{Berger2010a} has stated that $>26$ magnitude host galaxies can be explained by typical SGRB hosts ($L\approx 0.1-1 L^*$) at $z>2$. \citet{Tunnicliffe2014} also found that faint hosts at $z<1$ are ruled out but that higher redshifts are not excluded by the current limits on coincident hosts. In this case ($z>1$), the observationally hostless bursts would be well constrained to exist within or nearby to $5r_e$ (See Fig. \ref{obshostlesslimitsboth}). We further explore these possibilities in future work.

\section*{Acknowledgements}
The authors thank the referee for their quick response and helpful comments that improved the manuscript. The authors would also like to acknowledge Kenta Hotokezaka, Ehud Nakar, Alexander van der Horst, and Sylvain Guiriec for helpful discussions, and Eleonora Troja for useful comments that improved the manuscript. The authors likewise acknowledge Phillip F. Hopkins for discussions involving the gas density profile of galaxies. The work of BO was supported in part by the National Aeronautics and Space Administration through grants NNX16AB66G, NNX17AB18G, and 80NSSC20K0389. The research of PB was funded by the Gordon and Betty Moore Foundation through Grant GBMF5076. This work made use of data supplied by the UK \textit{Swift} Science Data Centre at the University of Leicester. The computations were performed on the George Washington University (GWU) Colonial One and Pegasus computer clusters.








\appendix

\section{Observationally Hostless SGRBs}
\label{Appendix: obs hostless}
Constraints on density as applied to five SGRBs identified as observationally hostless in the literature (see also \S \ref{sec_hostlesslimits}):

\textit{GRB 061201} -- For GRB 061201, \citet{Berger2010a} identified that a star forming late-type galaxy at $z=0.111$ has the lowest probability of chance coincidence; the galaxy is at a projected physical offset of $32.4$ kpc. Using deeper HST observations, \citet{FongBerger2013} identify galaxies at offsets of $16.3\arcsec$ and $1.8\arcsec$ with similar chance probabilities $P_{cc}\approx 0.07$. The host at an offset of $16.3\arcsec$ is the $z=0.111$ host identified by \citet{Berger2010a}. The equivalent chance probabilities for the two potential hosts leads to a observationally hostless classification. We infer a lower limit on circumburst density of $n_{\textrm{min}}=1.1\times 10^{-3}\, \textrm{cm}^{-3}$ for $z=0.111$ \citep[see also][]{Stratta2007}. Using the host galaxy magnitude, $m_{F160W}=18.63\pm0.01$ \citep{Fong2013}, we estimate a stellar mass of $M_*=(1.3^{+1.3}_{-0.7})\times 10^9 M_\odot$, which is at the low end of observed SGRB host masses. This estimate was obtained using stellar mass-absolute magnitude ($M_*-\mathcal{M}_{F160W}$) relations we derived using CANDELS UDS \& GOODS-S galaxies \citep{Santini2015}. For this mass, we obtain a density at offset, $n_\textrm{offset}=2\times 10^{-4}$ cm$^{-3}$, which is an order of magnitude lower than our lower limit. We find that the density at 32.4 kpc is only marginally changed by varying the gas fraction, as the inner gas profile is exponential (see \S \ref{sec: gas density profile}). These lines of evidence lead us to conclude that the $z=0.111$ late-type galaxy is an unlikely host for GRB 061201 if the gas density profile adopted in this work accurately resembles this galaxy's profile.

\par
\textit{GRB 070809} -- \citet{Berger2010a} identified that the galaxy with the lowest $P_{cc}$ is an early-type galaxy at $z=0.473$ with a projected physical offset of 34.8 kpc ($5.7\arcsec$). \citet{FongBerger2013} backs up this claim with a $P_{cc}=6\times10^{-3}$, which is a factor of three lower than that calculated by \citet{Berger2010a}. \citet{Zevin2019} investigated the required natal kick to produce the observed offset of 9.25$R_e$ by evolving both the orbit and galactic potential. They assume the host galaxy is the lowest $P_{cc}$ galaxy in the field, which has stellar mass $\log(M_*/M_\odot)=11.27$ and effective radius $R_e=3.59$ kpc. In order to produce the observed physical offset, \citet{Zevin2019} found that systemic kick velocities (combination of natal kick and mass ejecta) exceeding 190 km/s were required at the 90\% level. We estimate a lower limit on circumburst density $n_{\textrm{min}}=1.1\times 10^{-3}\, \textrm{cm}^{-3}$ for $z=0.473$ and $\varepsilon_B=10^{-2}$. Furthermore, we calculate a gas density at the physical offset (34.8 kpc) of $n_\textrm{offset}\approx 1.6\times 10^{-3}$ cm$^{-3}$ for this host galaxy mass and redshift, which is consistent with our lower limit.

\par
\textit{GRB 091109B} -- GRB 091109B has two potential hosts with $P_{cc}$ = 0.09 and 0.1, but neither have low enough chance probability to yield a strong host association \citep{Tunnicliffe2014}. HST observations by \citet{FongBerger2013} have a diffraction spike which limits their ability to identify a coincident host in the observations. The potential host with the lowest chance probability in their observations is located at $11.7\arcsec$ from the GRB position with $P_{cc}\approx 0.08$. We estimate a lower limit on circumburst density $n_{\textrm{min}}(z_\textrm{min})=1.7\times 10^{-5}\, \textrm{cm}^{-3}$. $n_\textrm{min}$ is higher than the predicted density at the GRB offset \cite[3\arcsec,][]{Tunnicliffe2014} for $z>0.7$. It is very plausible this GRB is not physically hostless given the average redshift of our sample, $\langle z\rangle= 0.84$. 
\par
\textit{GRB 110112A} -- This GRB has no convincing host associations. \citet{Fong2013} have stated it likely resided in a faint host. GRB 110112A differs from other observationally hostless GRBs in that it has no potential hosts with $P_{cc}<0.3$. The two most probable hosts from \citet{Fong2013} have $P_{cc}=0.43$ and $0.54$, whereas from \citet{Tunnicliffe2014} the three potential hosts have $P_{cc}=0.34$, $0.42$, and $0.50$. We calculate a lower limit $n_{\textrm{min}}(z_\textrm{min})=1.4\times 10^{-3}\, \textrm{cm}^{-3}$. More generally, the lower limit on density is inconsistent with being physically hostless for $z>0.1$ \citep[see also][]{Fong2013}. This implies that the lowest $P_{cc}$ host galaxy ($4.8\arcsec$) is an unlikely host, and we concur with \citet{Fong2013} that the BNS system likely resided in a faint host galaxy.
\par
\textit{GRB 111020A} -- GRB 111020A is in a crowded field and has multiple potential hosts with low probabilities of chance coincidence ($P_{cc}=0.007-0.09$), which complicates a convincing host identification \citep{Tunnicliffe2014}. The object with a chance probability of $P_{cc}=0.007$ could not be conclusively determined to be a galaxy instead of a faint star, and this prevented the strong host association \citep{Tunnicliffe2014}. We determine a lower limit of $n_{\textrm{min}}(z_\textrm{min})=1.3\times 10^{-3}\, \textrm{cm}^{-3}$, which is higher than the predicted density at the observed offset \citep[3\arcsec][]{Tunnicliffe2014} for $z>0.4$. This implies the GRB is not physically hostless. Therefore, we suggest a faint host at moderate to high redshifts (e.g., $z>2$).
\par
Additionally there are three observationally hostless bursts without quality X-ray observations that we exclude from of our sample. We do not set limits on circumburst density for these SGRBs (see \S \ref{sec_hostlesslimits}), but discuss the observations of potential hosts. For GRB 080503A, a galaxy located $0.8\arcsec$ from the afterglow position with magnitude $m_{F606W}=27.3\pm 0.2\, \textrm{mag (AB)}$ has the lowest probability of chance coincidence \citep{Perley2009}. This is an example of a dim host that would have been missed by the follow-up observations of the other observationally hostless bursts. A strong association with this host is not made because of another 4 faint galaxies within $2\arcsec$ that have $P_{cc}\approx 0.1-0.2$ \citep{Berger2010a}. Of the 5 observationally hostless bursts considered in \citet{Berger2010a} with optical afterglows (i.e., sub-arcsecond localization) and no coincident host galaxy, only GRB 080503A has lowest $P_{cc}$ for the nearest galaxy. \citet{FongBerger2013} identified the same most probable galaxy, but with a lower $P_{cc}$ by a factor of two. For a potential host galaxy of GRB 090305A, \citet{Tunnicliffe2014} measured a value of $P_{cc}=0.09$, but this value is not low enough to fall in their $P_{cc}\lesssim0.01$ category of strong host association. But, \citet{FongBerger2013} identified a potential host at an offset of $0.43\arcsec$ with chance probability $P_{cc}=7\times10^{-3}$. This galaxy was not previously identified in ground based observations from \citet{Berger2010a}. In the case of GRB 090515, the galaxy with the lowest $P_{cc}$ is an early-type galaxy with stellar mass $\log(M_*/M_\odot)=11.2$ at $z=0.403$ with a projected physical offset of 75 kpc ($14\arcsec$) from the GRB localization \citep{Berger2010a,Zevin2019}. \citet{Zevin2019} found that to produce the observed offset of 75 kpc (17.7$R_e$) systemic kick velocities $>270$ km/s (90\% CI) were required. There are 2 additional galaxies nearby the localization of GRB 090515 that prevent a strong host association due to their similar chance probabilities \citep{Berger2010a}. Using HST observations \citet{FongBerger2013}, lowered the chance probability by a factor of two ($P_{cc}\approx 0.05$) for the same most probable host identified by \citet{Berger2010a}. 

\section{The short gamma-ray burst sample}

In Table \ref{Table_SGRB_DATA}, we present the 52 SGRBs used in this work, of which 19 (36\%) have known redshifts. See \S \ref{subsec: SGRBs Sample} for details on sample selection.

\begin{table*}
\centering
\caption{We present the redshift $z$, BAT fluence $\phi_{\gamma,-7}=\phi_\gamma/10^{-7}$ erg/cm$^{2}$, upper limit on deceleration time $t_o$, unabsorbed flux $F_{X,o,-10}=F_{X,o}/10^{-10}$ erg/cm$^{2}$/s at $t_o$, and the bolometric correction factor $k_{\textrm{bol}}$ for SGRBs in our sample.}
\begin{tabular}{|c|c|c|c|c|c|}
\hline
\hline
GRB & $z$  & $\phi_{\gamma,-7}$ (erg/cm$^2$)  & $t_o$ (s) & $F_{X,o,-10}$ (erg/cm$^2$/s) & $k_{\textrm{bol}}$  \\
\hline
\hline
051221A & 0.546 & $12.0\pm0.4$  & 95$\pm3$ & $2.9\pm0.6$ & 7.1$^{+3.0}_{-2.3}$  \\[0.5mm]
060801 & 1.131 & $0.8\pm0.1$ & 118$^{+5}_{-7}$ & $2.3\pm0.52$ & 7.1$^{+4.9}_{-2.9}$   \\[0.5mm]
061006 & 0.438 & $14.2^{+1.4}_{-1.3}$ & 168$^{+22}_{-13}$ & $0.38\pm0.09$ & 12.4$_{-5.3}^{+8.2}$   \\[0.5mm]
070429B & 0.902 & $0.63\pm0.09$ & 520$_{-270}^{+350}$ & $0.014\pm 0.001$ & 7.3$_{-2.0}^{+2.6}$   \\[0.5mm]
070714B & 0.923 & $7.2\pm0.09$ & 68.6$\pm0.9$ & 12$_{-1.5}^{+2.0}$ & 20.8$_{-13.2}^{+23.4}$   \\[0.5mm]
070724A & 0.457 & $0.34\pm0.08$ & 336$_{-129}^{+591}$ & $0.23\pm0.04$ & 2.4$_{-0.07}^{+0.07}$   \\[0.5mm]
071227 & 0.381 & 6.0$\pm2.0$ & 506$_{-131}^{+303}$ & 0.04$\pm0.01$ & 12.2$_{-5.8}^{+11.05}$   \\[0.5mm] 
090426A & 2.609 & 1.8$_{-0.2}^{+0.3}$ & 255$_{-22}^{+23}$ & 0.32$\pm0.07$ & 2.0$_{-0.2}^{+0.1}$  \\[0.5mm] 
090510 & 0.903& $4.08\pm0.07$& 100$\pm2$ & 3.9$\pm0.8$ & 36.0$_{-2.4}^{+2.6}$   \\[0.5mm] 
100625A & 0.45 & $8.8\pm3.3$ & 175$_{-23}^{+30}$ & 2.7$\pm0.06$ & 7.5$_{-2.3}^{+3.7}$\\[0.5mm] 
100724A & 1.288 & 1.6$\pm0.2$ & 79$\pm4$ & 1.4$\pm0.4$ & 6.5$_{-2.6}^{+4.3}$   \\[0.5mm] 
101219A & 0.718 & 4.3$\pm0.25$ & 83$_{-6}^{+7}$ & 3.5$\pm0.8$ & 8.6$_{-2.7}^{+4.1}$   \\[0.5mm]
111117A & 2.211 & 1.4$\pm0.2$ & 108$_{-18}^{+19}$ & 1.04$\pm0.13$ & 4.5$_{-1.5}^{+2.3}$   \\[0.5mm]
120804 & 1.3 & 8.8$\pm0.5$ & 322$_{-11}^{+9}$ & 3.0$\pm0.7$ & 6.5$_{-2.6}^{+4.3}$   \\[0.5mm]
130603B & 0.356 & 6.3$\pm0.3$ & 49$\pm5$ & 2.1$\pm0.4$ & 12.5$_{-6.0}^{+11.4}$  \\[0.5mm]
140903A & 0.351 & 1.4$\pm0.1$ & 196$\pm61$ & 0.25$\pm0.05$ & 12.6$_{-6.0}^{+11.5}$   \\[0.5mm]
141212A & 0.596 & 0.73$\pm0.12$ & 180$_{-100}^{+320}$ & 0.017$\pm0.005$ & 10.1$_{-4.6}^{+8.4}$  \\[0.5mm]
150423A & 1.4 & 0.7$\pm0.1$ & 109$_{-30}^{+32}$ & 0.16$\pm0.04$ & 6.20$_{-2.4}^{+3.0}$  \\[0.5mm]
160821B & 0.16 & 1.2$\pm0.01$ & 379$_{-42}^{+116}$ & 0.11$\pm 0.02$ & 15.3$_{-7.7}^{+15.4}$   \\
\hline
\hline
060313 & -- & 11.2$\pm0.5$ & 99$_{-2}^{+3}$ & 2.6$\pm0.8$  & 18.7$^{+20.4}_{-9.7}$  \\[0.5mm]
061201 & -- & 3.3$\pm0.3$ & 107$_{-9}^{+8}$ & 2.1$\pm0.4$ &18.7$^{+20.4}_{-9.7}$   \\[0.5mm]
070809 & -- & 1.0$\pm0.2$ & 126$_{-46}^{+100}$ & 0.16$\pm0.04$ & 18.7$^{+20.4}_{-9.7}$  \\[0.5mm]
080123 & -- & 2.8$\pm1.7$ & 1200$\pm500$ & 0.015$\pm0.003$ & 18.7$^{+20.4}_{-9.7}$  \\[0.5mm]
080426 & -- & 3.7$\pm0.2$ & 269$^{+37}_{-39}$ & 0.45$\pm0.1$ & 18.7$^{+20.4}_{-9.7}$  \\[0.5mm]
080702A & -- & 0.3$\pm0.1$ & 126$^{+45}_{-50}$ & 0.3$\pm0.06$ & 18.7$^{+20.4}_{-9.7}$  \\[0.5mm]
080905A & -- & 1.4$\pm0.2$ & 1053$^{+193}_{-132}$ & 0.014$\pm0.003$ & 18.7$^{+20.4}_{-9.7}$  \\[0.5mm]
080919 & -- & 0.72$\pm0.1$ & 81$^{+4}_{-7}$ & 1.5$\pm0.3$ & 18.7$^{+20.4}_{-9.7}$  \\[0.5mm]
081024A & -- & 1.2$\pm0.16$ & 78$\pm2$ & 26$\pm4$ & 18.7$^{+20.4}_{-9.7}$   \\[0.5mm]
081226A & -- & 1.0$\pm0.2$ & 179$^{+80}_{-70}$ & 0.6$\pm0.14$ & 18.7$^{+20.4}_{-9.7}$  \\[0.5mm]
090621B & -- & 0.7$\pm0.1$ & 336$^{+500}_{-250}$ & 0.04$\pm0.01$ &18.7$^{+20.4}_{-9.7}$    \\[0.5mm]
091109B & -- & 2.0$\pm0.2$ & 155$^{+115}_{-70}$ & 0.1$\pm0.02$ & 18.7$^{+20.4}_{-9.7}$   \\[0.5mm]
110112A &--  & 0.3$\pm0.1$ & 114$\pm31$ & 0.2$\pm0.04$ & 18.7$^{+20.4}_{-9.7}$  \\[0.5mm]
111020A &--  & 0.7$\pm0.1$ & 124$^{+48}_{-40}$ & 0.4$\pm0.1$ & 18.7$^{+20.4}_{-9.7}$  \\[0.5mm]
111121A & -- & 22$\pm1.5$ & 66$\pm0.3$ & 69$^{+8}_{-9}$  &18.7$^{+20.4}_{-9.7}$     \\[0.5mm]
120305A & -- & 2.0$\pm0.1$ & 146$\pm3$ & 10.7$\pm2.4$ & 18.7$^{+20.4}_{-9.7}$   \\[0.5mm]
120630A & -- & 0.6$\pm0.2$ & 175$^{+120}_{-80}$ & 0.06$\pm0.01$ & 18.7$^{+20.4}_{-9.7}$  \\[0.5mm]
121226A & -- & 1.4$\pm0.2$ & 207$\pm20$ & 1.6$\pm0.4$ &18.7$^{+20.4}_{-9.7}$    \\[0.5mm]
130515A & -- & 1.5$\pm0.2$ & 377$^{+420}_{-130}$ & 0.015$\pm0.004$ & 18.7$^{+20.4}_{-9.7}$   \\[0.5mm]
130912A & -- & 1.7$\pm0.2$ & 155$^{+15}_{-13}$ & 2.4$\pm0.5$ & 18.7$^{+20.4}_{-9.7}$   \\[0.5mm]
140129B &--  & 0.7$\pm0.1$ & 355$\pm10$ & 0.9$\pm0.2$ & 18.7$^{+20.4}_{-9.7}$   \\[0.5mm]
140930B &--  & 4.2$\pm0.4$ & 187$\pm4$ & 4.0$\pm0.5$ & 18.7$^{+20.4}_{-9.7}$   \\[0.5mm]
150301A & -- & 0.7$\pm0.1$ & 52$\pm1$ & 59$\pm9$ & 18.7$^{+20.4}_{-9.7}$  \\[0.5mm]
151127A & -- & 0.23$\pm0.06$ & 174$^{+94}_{-77}$ & 0.08$\pm0.02$ & 18.7$^{+20.4}_{-9.7}$   \\[0.5mm]
160408A & -- & 1.6$\pm0.2$ & 850$^{+570}_{-200}$ & 0.015$\pm0.003$ &18.7$^{+20.4}_{-9.7}$    \\[0.5mm]
160411A &--  & 0.8$\pm0.2$ & 3350$^{+270}_{-210}$& 0.018$\pm0.004$ & 18.7$^{+20.4}_{-9.7}$  \\[0.5mm]
160525B & -- & 0.32$\pm0.08$ & 73$\pm3$ & 1.3$\pm0.3$ & 18.7$^{+20.4}_{-9.7}$   \\[0.5mm]
160601A & -- & 0.7$\pm0.1$ & 270$^{+420}_{-190}$ & 0.012$\pm0.003$ & 18.7$^{+20.4}_{-9.7}$   \\[0.5mm]
160927A & -- & 1.4 $\pm0.2$& 109$\pm21$ & 0.23$\pm0.5$ &18.7$^{+20.4}_{-9.7}$   \\[0.5mm]
170127B & -- & 1.0$\pm0.2$ & 130$^{+50}_{-30}$ & 0.6$\pm0.1$ &18.7$^{+20.4}_{-9.7}$   \\[0.5mm]
170428A & -- & 2.8$\pm0.2$ & 800$^{+140}_{-100}$ & 0.1$\pm0.03$ &18.7$^{+20.4}_{-9.7}$    \\[0.5mm]
180402A & -- & 1.4$\pm0.2$ & 87$^{+2}_{-3}$ & 5$\pm1$ & 18.7$^{+20.4}_{-9.7}$  \\[0.5mm]
181123B & -- & 1.2$\pm0.2$ & 125$^{+20}_{-30}$ & 0.32$\pm0.07$ &18.7$^{+20.4}_{-9.7}$   \\
\hline
\end{tabular}
\label{Table_SGRB_DATA}
\end{table*}

\section{11-Hour Flux Correlation: Monte Carlo Simulations}
\label{Appendix: Monte Carlo}
 Here we describe the Monte Carlo simulations of \S \ref{sec: 11-hour flux corr} used to constrain the width of the SGRB circumburst density distribution. As discussed in \S \ref{sec: lower limits methods sec}, we consider synchrotron radiation from electron's accelerated in the forward shock, including IC corrections, to calculate the X-ray flux (0.3-10 keV) at 11-hours. The gamma-ray luminosity is sampled using the luminosity functions from G16 (their case $a$) and WP15. We adopt a rest frame duration $\langle T_{90}\rangle=0.2$ s in order to calculate the fluence. The redshift is simulated according to a convolution of the star formation rate density from \citet{Moster2013} and the delay time distribution from \citet{Beniamini2019}. We also adopt a scatter in $E_{p,\textrm{obs}}$, which affects the bolometric correction, using the standard deviation observed by \citet{Nava2011}. We convert the distribution of $E_{p,\textrm{obs}}$ to the source frame using $E_{p,\textrm{source}}=E_{p,\textrm{obs}}(1+\langle z\rangle)$ as previously described in \S \ref{subsec: SGRBs Sample}. We apply lognormal distributions for the parameters $\varepsilon_e$ and $\varepsilon_B$. We adopt $\langle \log \varepsilon_e\rangle=-1$ and $\sigma_{\log \varepsilon_e}=0.3$ as previous works have shown that $\varepsilon_e$ has a narrow distribution \citep{Nava2014,BeniaminiVanDerHorst}. For $\varepsilon_B$, we test the effect of varying $\langle \log \varepsilon_B\rangle$ with fixed $\sigma_{\log \varepsilon_B}=1$ as used by \citet{Beniamini2016corr}. This is a typical scatter in $\varepsilon_B$ found from GRB afterglow modeling \citep{Santana2014,Zhang2015}. The parameters $\varepsilon_\gamma=0.15$ and $p=2.2$ are fixed as in \S \ref{sec: lower limits methods sec}. This is a valid assumption as recent works have shown that $\varepsilon_\gamma$ must have a narrow distribution \citep{Nava2014,Beniamini2016corr,BeniaminiVanDerHorst}. We apply detection thresholds on the flux and fluence of $F_{X,11}\gtrsim 10^{-15}$ erg/cm$^{2}$/s and $\phi_\gamma\gtrsim 10^{-8}$ erg/cm$^{2}$ which were determined based on the lowest value of SGRBs with measurements of 11-hour X-ray flux used in this work. We draw $N=10^5$ realizations of this Monte Carlo simulation in order to determine the density distribution required to reproduce the observed correlation (e.g., standard deviation in best fit). In addition, we track the fraction of detected SGRBs for which electrons radiating synchrotron at $\nu_X\approx 1$ keV are in the slow cooling (SC) regime. This is important because the flux from electrons in the slow cooling regime is $\propto n^{1/2}$, whereas for $n>n_{\rm crit}$, the flux is in the fast cooling regime, and therefore independent of density (\ref{eqn: 11-hour flux FC}). The results of these simulations are described in \S \ref{sec: 11-hour flux corr} and tabulated in Table \ref{fig: 11-hr corr}.


\bsp	
\label{lastpage}
\end{document}